\documentclass{aa}

\usepackage{graphicx}
\usepackage{txfonts}
\newcommand{\Myr}{\, \rm{Myr} }

\newcommand{\Mpc}{\, \rm{Mpc} }

\newcommand{\Msun}{\, \rm{M_{\odot}} }

\begin{document}

   \title{The \textsc{LoReLi} database: 21 cm signal inference with 3D radiative hydrodynamics simulations}

   \author{R.Meriot
          \inst{1}
          \and
          B.Semelin\inst{1}
          }

   \institute{ Observatoire de Paris, PSL Research University,  Sorbonne Université, CNRS, LERMA, 75014 Paris, France}

 
  \abstract
   {The Square Kilometer array is expected to measure the 21cm signal from the Epoch of Reionization (EoR) in the coming decade, and its pathfinders may provide a statistical detection even earlier. The currently reported upper limits provide tentative constraints on the astrophysical parameters of the models of the EoR.
   In order to interpret such data with 3D radiative hydrodynamics simulations using Bayesian inference, we present the latest developments of the \textsc{Licorice} code. Relying on an implementation of the halo conditional mass function to account for unresolved star formation, this code now allows accurate simulations of the EoR at $256^3$ resolution. We use this version of \textsc{Licorice} to produce the first iteration of \textsc{LoReLi}, a public dataset now containing hundreds of 21cm signals computed from radiative hydrodynamics simulations. We train a neural network on \textsc{LoReLi} to provide a fast emulator of the \textsc{Licorice} power spectra, \textsc{LorEMU}, which has $\sim 5\%$ rms error relative to the simulated signals. \textsc{LorEMU} is used in a Markov Chain Monte Carlo framework to perform Bayesian inference, first on a mock observation composed of a simulated signal and thermal noise corresponding to 100h observations with the SKA. We then apply our inference pipeline to the latest measurements from the HERA interferometer. We report constraints on the X-ray emissivity, and confirm that cold reionization scenarios are unlikely to accurately represent our Universe.}



   \maketitle
%

\section{Introduction}

The first billion years of the evolution of the Universe constitute a key epoch in its development on large scales. Following the hierarchical theory of structure formation and the physics regulating star formation, the gas contained in massive halos cools down, subsequently fragmenting, collapsing, and giving birth to the first stars. This marks the end of the "dark ages" in the history of the Universe. There are several processes responsible for this cooling. The atomic hydrogen cooling channel allows significant cooling down to $\sim 10^4$ K, that is below the virial temperature of halos with masses of greater than  $M_{min} \approx 10^{8} \Msun$, allowing gravothermal collapse and star formation. Molecular hydrogen cooling brings this minimal mass for star formation down to  $M_{min} \approx 10^{5} \Msun$, in so-called "mini halos". 
The emergence of the first stars and galaxies, in a period called Cosmic Dawn (CD, $z\sim15-30$), greatly impacted the neutral gas in the intergalactic medium (IGM). Ultraviolet radiation from the first generation of sources ionized their local environment, while the still neutral IGM, farther away from the sources, became heated by X-ray emissions. During the Epoch of Reionization (EoR, $z\sim6-15$), the ionized regions quickly grew until they covered the entire IGM.       
The 21 cm line of neutral hydrogen is emitted by the neutral hydrogen in the IGM and its intensity fluctuations are shaped by the astrophysical properties of the first sources, making it an extremely promising source of information regarding the first billion years of the evolution of the Universe. \cite{Furlanetto2006a} provide a review of this topic.


While valuable, the detection of this signal remains an observational challenge. Foreground emissions, which are roughly 10 000 times stronger than the expected intensity of the 21 cm signal, must be subtracted from observed data or avoided in order to extract the cosmological signal. The subtraction requires the very good handling of radio frequency interference, a deep and exact sky model, and accurate direction-dependent calibration \citep[see e.g.][]{Mertens2020}. Many instrumental programs are tackling this challenge, the most ambitious being the Square Kilometer Array (SKA). The unprecedented sensitivity of the SKA will enable us to probe this epoch with sufficient signal-to-noise ratio to build a full tomography of the signal between $z\sim 27$ and $z\sim6$, providing a comprehensive view of how reionization unfolded. The observation of the 21 cm signal by the SKA will start by the end of this decade, but several instruments are already trying to measure summary statistics of the signal. A first detection of the global (sky-averaged) signal at $z\sim16$ was announced by \cite{Bowmana}, who used the Experiment to Detect the Global EoR Signature (EDGES) instrument. The features of this detection, in particular the strong intensity of the signal in absorption, are not compatible with standard models of reionization and this result has not been confirmed by other global signal experiments, such as SARAS-3 \citep{Singh2021}. Interferometers such as the Low-Frequency Array (LOFAR), the Murchison Widefield Array (MWA), the New Extension in Nançay Upgrading LOFAR  (NenuFAR), and the Hydrogen Epoch of Reionization Array (HERA) aim to measure the power spectrum of the 21 cm signal. While no detection has yet been claimed, upper limits on the spectrum have been devised (e.g., \cite{ Mertens2020, Trott2020,TheHERACollaboration2022}; Munshi et al. (in prep.), and references therein). In the coming years, these instruments are expected to detect the signal or to produce upper limits that are informative enough to constrain nonexotic models of the EoR at various scales and redshifts. Given these observational prospects, in order to extract astrophysical information, the community must provide accurate models of the first billion years of evolution of the Universe, as well as inference methods linking raw observational data to the astrophysical processes encoded in the models.  

On the one hand, sustained effort has been put into developing theoretical and numerical approaches to model the EoR, ranging from fast semi-analytical codes  \citep[e.g.,][]{Mesinger2010, Santos2010, Cohen1915, Murray2020, Reis2021} to expensive 3D radiative transfer simulations \citep[e.g.,][]{Ciardi2003, Mellema2006, Baek2010, Semelin2017, Ocvirk2018, Lewis2021, Doussot2022}. Fast codes compute the dynamics of dark matter in the linear regime and typically assume a constant gas-to-dark-matter density ratio. On large scales (more than a few cMpc), this yields similar results to full numerical simulations that compute the nonlinear dynamics of gravity and hydrodynamics. However, star formation, a crucial process when modeling the EoR, occurs on small scales. This leads to very different modeling approaches. In high-resolution N-body simulations, individual dark matter halos are resolved.
Star formation within the halos can then be computed from the local properties of the gas (typically density and temperature), though the actual equations linking gas properties to star formation may differ between simulation codes. One-dimensional (spherically symmetric) radiative transfer simulations assign baryons to halos formed in pre-computed dark-matter-only simulations \citep{Krause2018, Ghara2018, Schaeffer2023} using simple recipes, and compute star formation from there. Semi-analytical codes usually do not resolve individual halos but estimate populations of halos and stars in a given region, through for instance the CMF formalism \citep{Lacey1993}. These different modeling approaches can potentially induce different features in the reionization field. The second critical step is the process through which UV and X-rays are accounted for. Fully numerical simulations typically include 3D radiative transfer through the use of M1 or various ray-tracing methods, while semi-analytical codes identify ionized and heated regions through simple (and easy to compute) photon budget arguments. A comprehensive evaluation of the impact of these different modeling approaches on the interpretation of the 21cm signal remains to be performed.  

On the other hand, applications of inference methods to the 21 cm signal have been under development in the last 10 years. The most widely employed technique is undoubtedly the Markov Chain Monte Carlo (MCMC) algorithm (see e.g., \cite{Greig2015,  Greig2018, Schmit2018}). In this approach, the posterior distribution of the parameters is explored in a guided random walk. This method is guaranteed to converge towards the true posterior probability (under some reasonable assumptions). However, it requires that an explicit likelihood can be formulated and it might require hundreds of thousands or millions of steps (and thus of realisations of the model) to produce a good estimate of the posterior, making its use only suited to fast numerical codes. This led the community to explore other inference techniques, such as training artificial neural networks to do the inference \citep{Shimabukuro2017, Doussot2019, Hortua, Neutsch2022}. While requiring fewer simulations in the training sample, convergence towards the true maximum likelihood or true posterior is in this case not guaranteed, as the training of the neural network can produce a systematic bias or a residual additional variance. Another noteworthy family of methods is called "simulation-based inference" or "likelihood-free inference" \citep{Alsing2017, Alsing2018, Zhao2022, Prelogovic2023}. These methods rely on the fact that a simulation is in and of itself a draw in the joint probability of the observable data and the model parameter, a joint probability that need not necessarily be explicitly writable. The joint probability distribution is then fitted using a neural network acting as a parametric function, allowing the straightforward computation of the posterior.

Due to computation time constraints, most inference methods have until now only been applied to signals generated using semi-numerical approaches, as they typically are a hundred thousand times faster than fully numerical simulations. Indeed, performing a single MCMC inference with the \textsc{Licorice} code, for example, running at high resolution would require $10^{11}$ CPU hours of computation, well beyond current capabilities. Performing inference with such a code at a reasonable cost therefore requires the  use of machine-learning-based approaches, which can reduce the number of simulations to a few thousand. However, even in this case, a decrease in resolution is necessary to reach an affordable computing cost. This decrease must be compensated for by a significant amount of subgrid modeling. This constitutes the approach at the heart of this work, where we use of the order of $10^3$ simulations to chart a 4D astrophysical parameters space with $256^3$ simulations in $300$ cMpc boxes. When then train neural network emulators for the power spectrum of the signal and use them to perform MCMC inference. 

In this paper, we present \textsc{LoReLi}, a collection of numerical simulations of the EoR using the \textsc{Licorice} code. In section  \ref{sec:Licoricecode}, we explain the recent modifications of the code that make such a database possible. In section \ref{sec:LoReLi}, we give details about the \textsc{LoReLi} database, and in section \ref{sec:inference}, we show the results of parameter inference applied to both mock data and the most recent data from the HERA collaboration \citep{Abdurashidova2022}, which we obtain using an emulator of the code trained on \textsc{LoReLi}.  

\section{The Licorice code}\label{sec:Licoricecode}


The simulations of the \textsc{LoReLi} dataset were run using the \rm{\textsc{Licorice}} simulation code \citep{Semelin2007, Baek2009, Semelin2015, Semelin2017}. 
 Here, we provide a brief overview of the code and details about recent developments. \textsc{Licorice} is an N-body simulation code dealing with baryons and dark matter particles.  Dynamics is solved using a TREE+SPH method \citep{Springel1998}. Radiative transfer in UV and X-ray continuum is coupled to dynamics through a Monte-Carlo scheme on an adaptive grid: photon packets are emitted in random directions by source particles (baryon particles with a nonzero stellar fraction), propagate at the speed of light, redshift, and deposit their energy in gas particles lying in the encountered cells depending on local density. They represent UV ionizing radiation and X-rays. "Hard" X-rays, after propagating over a distance larger than the size of the simulation box, are added to a uniform background.
 The ionization and temperature states of each gas particle are then updated several times per dynamical time step and the latter affect dynamics through the pressure and artificial viscosity terms in the SPH scheme \citep{Monaghan1992, Springel1998}. 

 \subsection{Star formation model}
 
 In previous iterations of \rm{\textsc{Licorice}}, gas particles with an SPH-calculated overdensity $\delta$ greater than a fixed threshold $\delta_{thresh}$ transform ---at each dynamical time step--- a fraction of their gas mass into stellar mass following a Schmidt-Kennicutt law of exponent one, using  

 \begin{equation}\label{eq:sf}
     df_{*} = (f_{coll} - f_{*})\dfrac{dt}{\tau}
 ,\end{equation}


\noindent where $f_{*}$ is the current stellar fraction of the particle, $f_{coll}  = 1 $ if $\delta>\delta_{thresh}$ and $f_{*}$ otherwise, $df_{*}$ is the newly created stellar fraction, $dt$ the duration of a time step, and $\tau$ is an astrophysical parameter of the simulation representing the typical gas-to-stars conversion timescale, thus controlling the star formation rate (SFR). However, this star formation numerical method is sensitive to an insufficient  resolution: failure to resolve the small-scale modes of the density field will smooth out the density peaks and prevent particles from reaching the density threshold for star formation, causing Eq. \eqref{eq:sf} to underestimate the SFR. Essentially, a simulation will lack the star formation of unresolved star-forming halos. The mass of the lightest of these halos is estimated to be  $\sim 10^8 \Msun$, as it corresponds to a virial temperature of $10^4$ K, the lowest temperature reached through the cooling of atomic hydrogen. In boxes of hundreds of cMpc, which is large enough to capture relevant large-scale fluctuations of the 21 cm signal, resolving such halos requires more than $4096^3$ particles and $10^{7\sim8}$ cpuh. This motivates a new implementation in the code of a subgrid model designed to provide a better estimate of star formation in unresolved halos, without affecting resolved ones.

\subsection{The conditional mass function subgrid model}

The chosen subgrid model relies on the conditional mass function (CMF) formalism \citep{Lacey1993} to estimate the halo mass functions of regions below the spatial resolution limit. In the CMF theory, the density of spheres of decreasing radii follows a random walk, and statistically estimating the radius at which the threshold $\delta_c$ is crossed for the first time gives the CMF of regions parameterized by their radius $R_0$ and overdensity $\delta_0$.

In extended Press-Schechter (EPS), the simplest formulation of the CMF theory, the fraction of the mass of a region (again parameterized by a radius $R_0$ and overdensity $\delta_0$) that lies in collapsed halos more massive than $M_{min}$ is given by 

\begin{equation}\label{eq:fcoll}
    f_{coll} = \rm{erfc}\left[ \frac{\delta_c-\delta_0}{\sqrt{2(\sigma(M_{min})^2 - \sigma_0^2 })}\right]
,\end{equation}

\noindent where $\sigma(M)$ is the density variance on mass scale $M$ and $\delta_c$ is the linear overdensity of collapse. The classical theory of structure formation predicts  $\delta_c \approx 1.68$. 

\cite{Sheth2001} devised a more accurate formalism (ST) to calculate the mass function. During the random walk, instead of crossing a constant threshold of size $\delta_c$, a "moving barrier"

\begin{equation}
    B(\sigma^2, z)  = \sqrt{a}\delta_c[1 + \beta(a\frac{\delta_c^2}{\sigma^2})^{-\alpha}]
\end{equation}

\noindent is adopted. This extends the EPS formalism from the physics of spherical collapse to that of ellipsoidal collapse. $\alpha  \approx 0.615, \beta  \approx 0.485$ are parameters whose values come from ellipsoidal dynamics and $a \approx 0.707$ is fitted to accurately model simulations.
\cite{Rubino-Martin2008, Tramonte2017} provided an expression for the CMF associated with ST:  

\begin{equation}\label{eq:st_cmf}
      n_c(M) =  -\sqrt{\frac{2}{\pi}}\frac{d\sigma}{dM} \frac{\rho_0}{M} \frac{|T(\sigma^2 | \sigma_0^2)|\sigma}{(\sigma^2- \sigma_0^2)^{3/2}} \rm{exp} \left[ -\frac{(B(\sigma^2, z) - \delta_0)^2}{2(\sigma^2 - \sigma_0^2)} \right]
,\end{equation}

\noindent  where 

\begin{equation}
    T(\sigma^2 | \sigma_0^2) = \sum^5_{n=0} \frac{(\sigma_0^2 - \sigma^2)}{n!} \frac{\partial^n[B(\sigma^2, z) - \delta_0]}{\partial(\sigma^2)^n}.
\end{equation}

\noindent  One can then easily calculate $M_{coll}$, the mass of the region contained in halos more massive than $M_{min}$ :

\begin{equation}\label{eq:cmf}
      M_{coll} = V_{region} \int_{M_{min}}^{M_{region}}  n_c(M)M\rm{dM}
  ,\end{equation}
\noindent  where $ V_{region}$ is the Lagrangian volume of the region and $M_{region}$ its mass. We then calculate the collapsed fraction as  

\begin{equation}\label{eq:fcoll2}
    f_{coll} =  \frac{M_{coll}}{M_{region}}.
\end{equation}


Both EPS and ST CMFs are implemented in the version of \textsc{Licorice} used in the present work.
In practice, we calculate $f_{coll}$ for each gas particle. To do so, we calculate the overdensity $\delta_0 = \frac{\rho_0 - \rho_m}{\rho_m}$  of the region they represent, where $\rho_0$ is the SPH-calculated density of the particle multiplied by the cosmological ratio $\Omega_m/\Omega_b$. The volume of the region is given by $V = \gamma\frac{4}{3}\pi R_0^3$, where $R_0$ is the SPH smoothing radius. While the formulation of the CMFs relies on top-hat filters, the SPH smoothing kernel is not a top-hat, and so the $\gamma$ parameter is tuned to represent the volume of the region of identical density smoothed with a top-hat kernel. Motivated by the fact that the integral of the SPH kernel reaches 99\% of its maximum value past $\sim 0.7 R_0$, we find  $\gamma = 0.4$ to be a reasonable choice, and indeed this choice is validated by the resolution study presented below.

We use the resulting value of $f_{coll}$ in Eq. \ref{eq:sf} instead of setting $f_{coll}$ depending on some density threshold. This is done either with Eq. \eqref{eq:fcoll} when using the EPS formalism, or with Eq. \eqref{eq:st_cmf}, \eqref{eq:cmf} and \eqref{eq:fcoll2} when using ST.

A final subtlety lies in the fact that in \textsc{Licorice}, all gas particles with a nonzero stellar fraction are considered sources of photons. However, this CMF formalism assigns a strictly positive $f_{coll}$ to all particles. This causes particles to behave in an unphysical manner, as they instantly turn into sources at the beginning of the simulation with infinitesimal stellar fractions. To avoid this, a method to stochastically assign $f_{coll}$ was adopted: when $df_{coll}$, the difference in $f_{coll}$ of a particle between one time step and the next, is smaller than $df_{coll,min} = \frac{M_{min}}{M_{region}}$, $df_{coll,min}$ is assigned to the particle with a probability $\frac{df_{coll}}{df_{coll,min}}$. This effectively prevents very weak, unphysical star formation at very high redshifts ($z \gtrsim 25-30$ depending on $M_{min}$, approximately when star formation starts in high-resolution simulations) while ensuring that the average star formation density (SFRD) remains identical to that of the nonstochastic model.

\begin{figure}
    \centering
    \includegraphics[scale = 0.36]{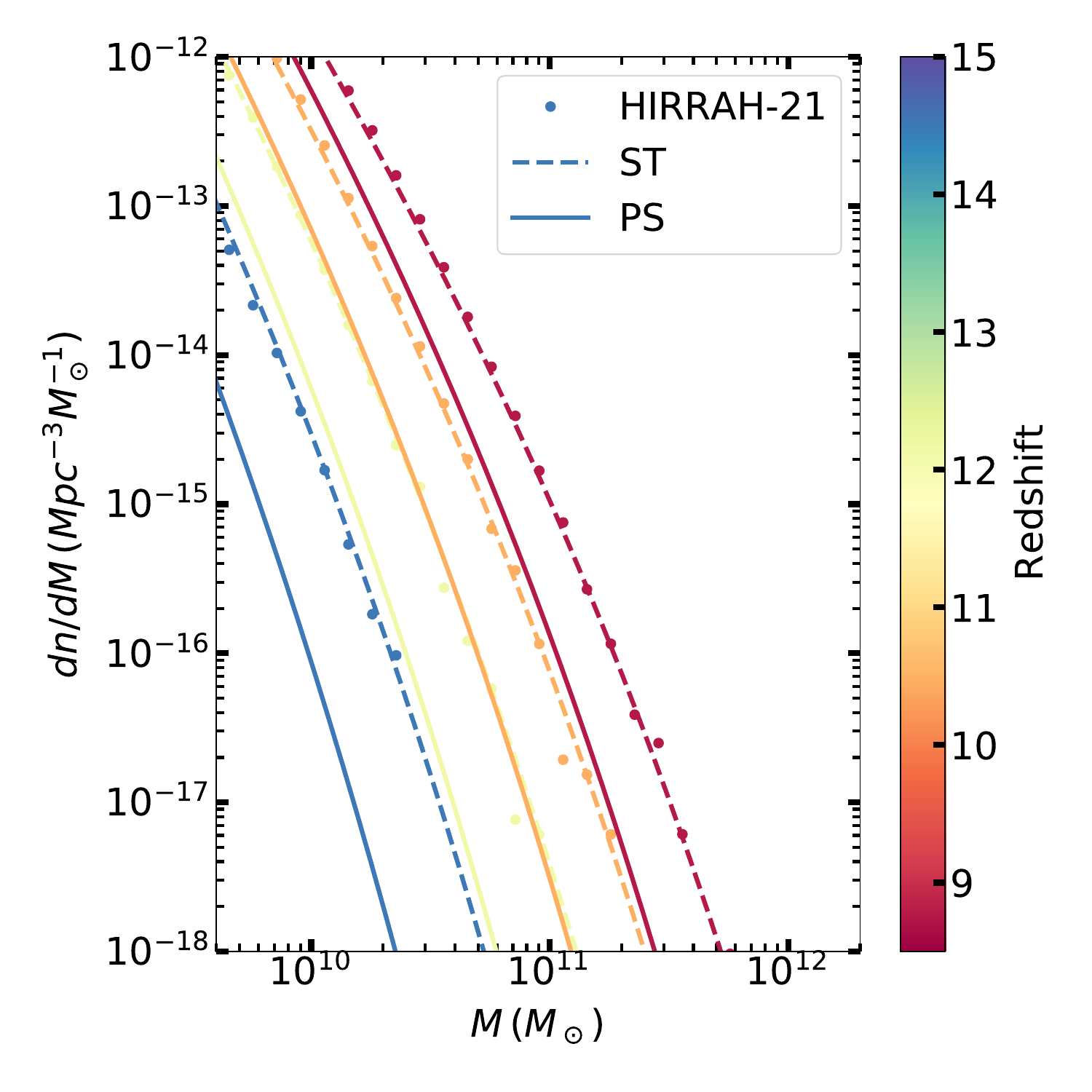}
    \caption{Evolution with redshift of the halo mass function of HIRRAH-21, calculated using a FoF algorithm. We also show the history of theoretical PS and ST HMFs for the same cosmology.
    }
    \label{fig:hmf}
\end{figure}

\begin{figure}
    \centering
    \includegraphics[scale = 0.36]{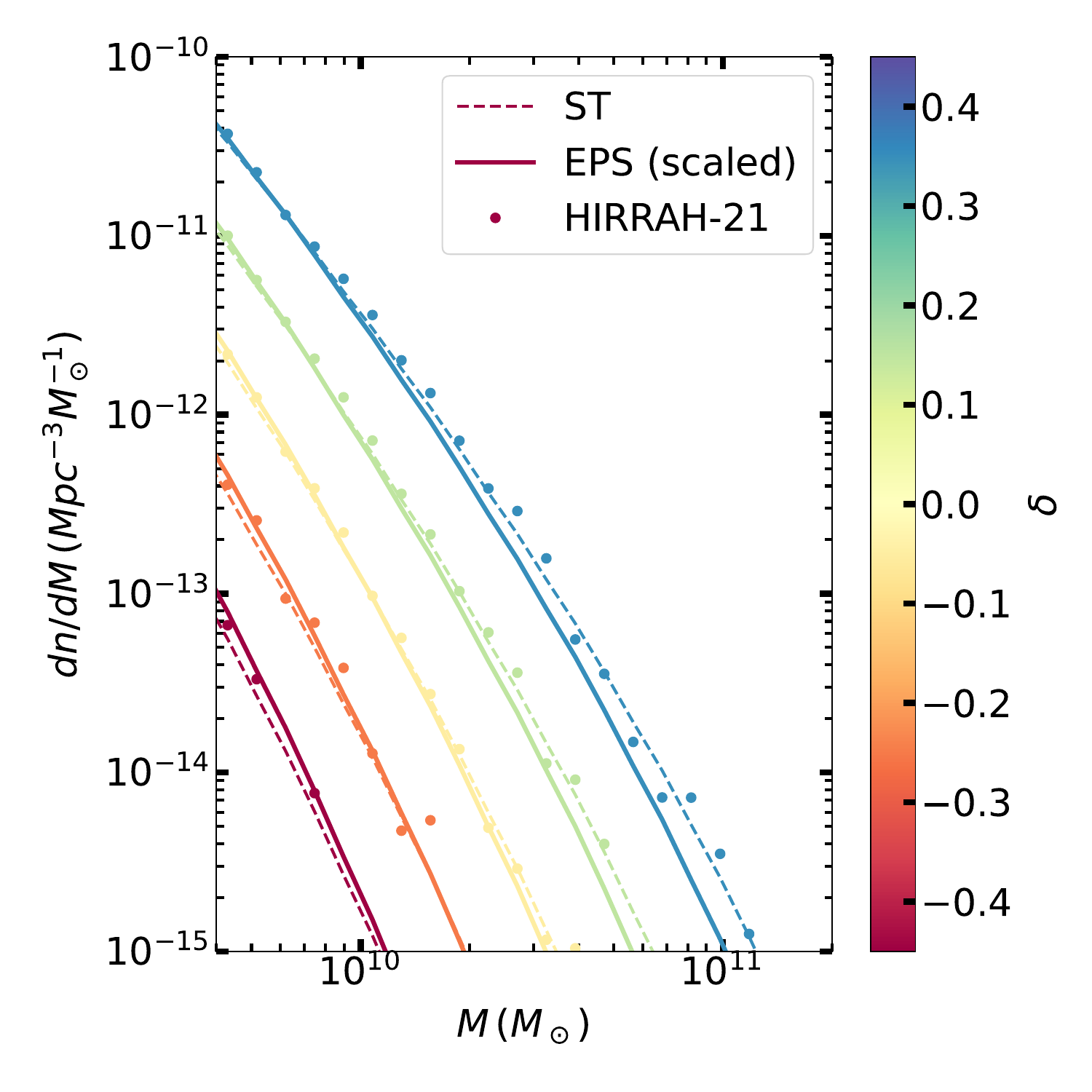}
    \caption{Evolution of the simulated CMF with the overdensity $\delta$ and of the theoretical CMFs as predicted by the ST and EPS models. In order to match the high-resolution simulation, the $\sigma$ in the EPS model has been scaled up by $5 \%$ on average.    }
    
    \label{fig:cmf_deltadep}
\end{figure}

Figure \ref{fig:hmf} shows how the HMF of HIRRAH-21, a high-resolution \textsc{Licorice} simulation ($2048^3$ particles, with no subgrid model) presented in \cite{Doussot2022}, compares with the PS and ST models at different redshifts.  The HMF of HIRRAH-21 was calculated by finding halos in the simulation using a Friends-of-Friends (FoF) algorithm. We find excellent agreement with the ST HMF, which lies a couple orders of magnitude above the PS HMF. This is not unexpected, given the findings of \cite{Reed2007}, and this difference in the HMFs necessarily appears in the CMFs. This justifies the implementation and use of the ST CMF. However, for practical reasons, in the presented and current version of the LoReLi database, we compute star formation using EPS and a different choice of an effective radius (10\% smaller, which implies a $\sim5\%$ larger $\sigma$) that produces a very similar SFRD. Future versions of the database will use the ST CMF. A comparison of the conditional mass functions of HIRRAH-21 at $z=9.48$ with the prediction of ST and EPS (with a $\sigma$ scaled up by 5\%) at different overdensities $\delta$ can be seen in Fig. \ref{fig:cmf_deltadep}. Good agreement between simulation and models is observed, with errors roughly within the fluctuations in the simulated CMF due to cosmic variance. The resulting $f_{coll}$ are within $\sim 30\%$ of each other.

\begin{figure}
    \centering
    \includegraphics[scale = 0.36]{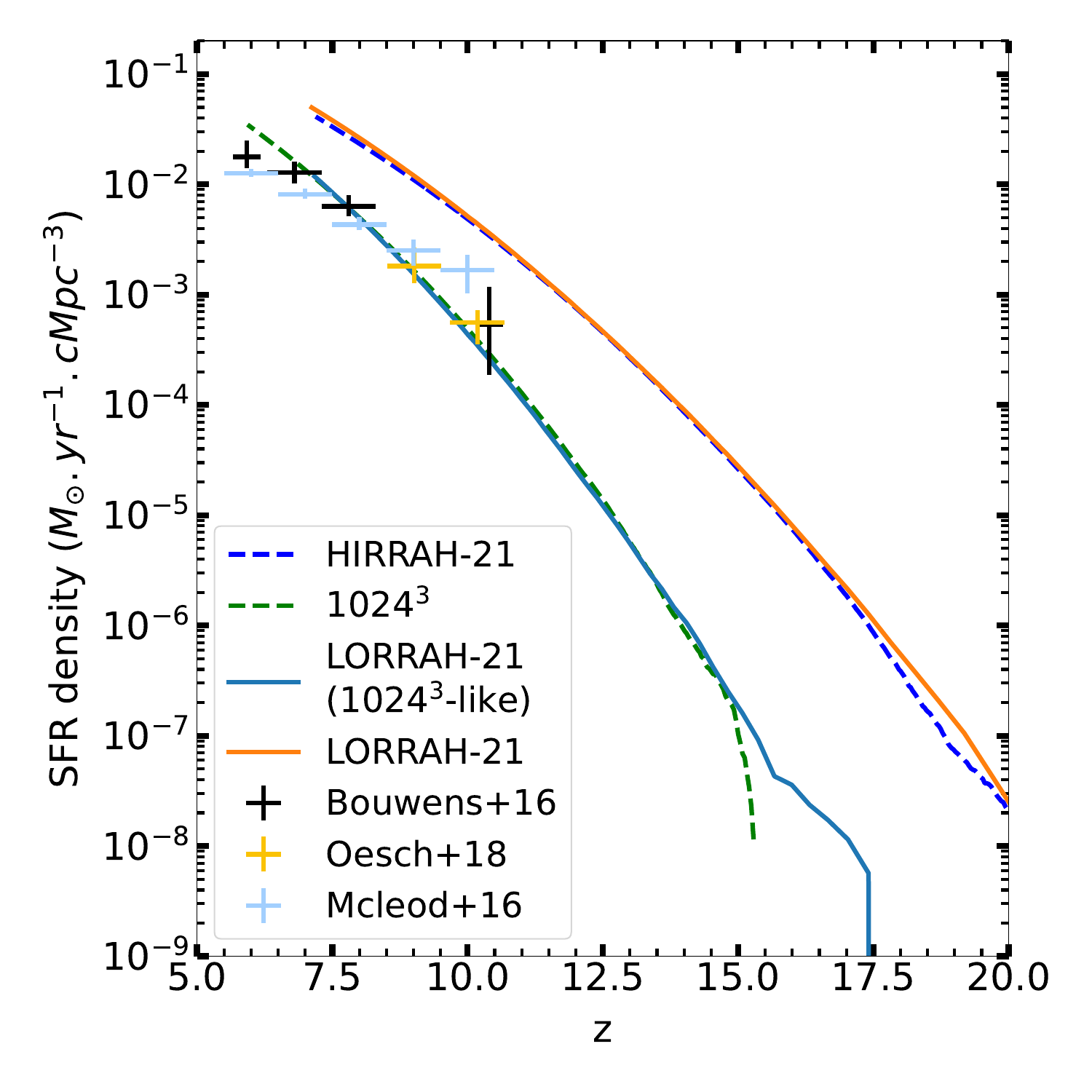}
    \caption{History of the SFR of HIRRAH-21 and LORRAH-21. As a reference, observational data from \cite{Bouwens2016, Oesch2018, McLeod2016} are included.  The SFRD of a $1024^3$ particles simulation is also plotted, and is consistent with that of a LORRAH-21 simulation using a $M_{min}$ parameter set at the value of the smallest resolved halos in the $1024^3$ simulation.  }
    \label{fig:sfr_testpred}
\end{figure}

However, the  connection between the simulated HMFs and star formation in the simulation is indirect, as star formation in \textsc{Licorice} does not necessarily occur in identified dark matter halos: only the local density of the gas appears in the star formation equation and not the properties of the dark matter that may or may not be present nearby. In order to evaluate the results of this subgrid model on star formation, we compare in Fig. \ref{fig:sfr_testpred} the star formation rate density (SFRD) of HIRRAH-21 and LORRAH-21, a low-resolution simulation ($256^3$ particles with the subgrid modeling of star formation). The $M_{min}$ parameter of the subgrid model was set at $4\times 10^9 \Msun$ to match the mass of the lightest halos identified by FoF in HIRRAH-21, and all other astrophysical and numerical parameters are identical in the two setups except for the dynamical time step: $0.5\Myr$ in HIRRAH-21 and $7\Myr$ in LORRAH-21. We observe very good agreement between the SFRDs of LORRAH-21 and HIRRAH-21 over the entire redshift range (with an RMS of the relative error of $0.16$ between $z\sim20$ and $z\sim7$, and of $0.09$ between $z\sim15$ and $z\sim7$). The largest mismatch ($\sim 30\%$) occurs at $z\sim18-20$, when the stochastic attribution of $f_{coll}$ is the most relevant. 
In order to test this subgrid model at a different resolution, Fig. \ref{fig:sfr_testpred} also shows the SFRD in a $1024^3$ simulation (from the 21SSD database \cite{Semelin2017}) with the same parameters as HIRRAH-21 (but a minimal mass of DM halos naturally set to  $3.2\times 10^{10} \Msun$) and in LORRAH-21, this time with $M_{min}$ set to the same value. In this case as well, excellent agreement is observed, implying that this CMF formalism is robust to resolution and can accurately describe the behavior of simulations of higher resolution.

\subsection{Two-phase model of gas particles}

\begin{figure}
    \centering
    \includegraphics[scale = 0.36]{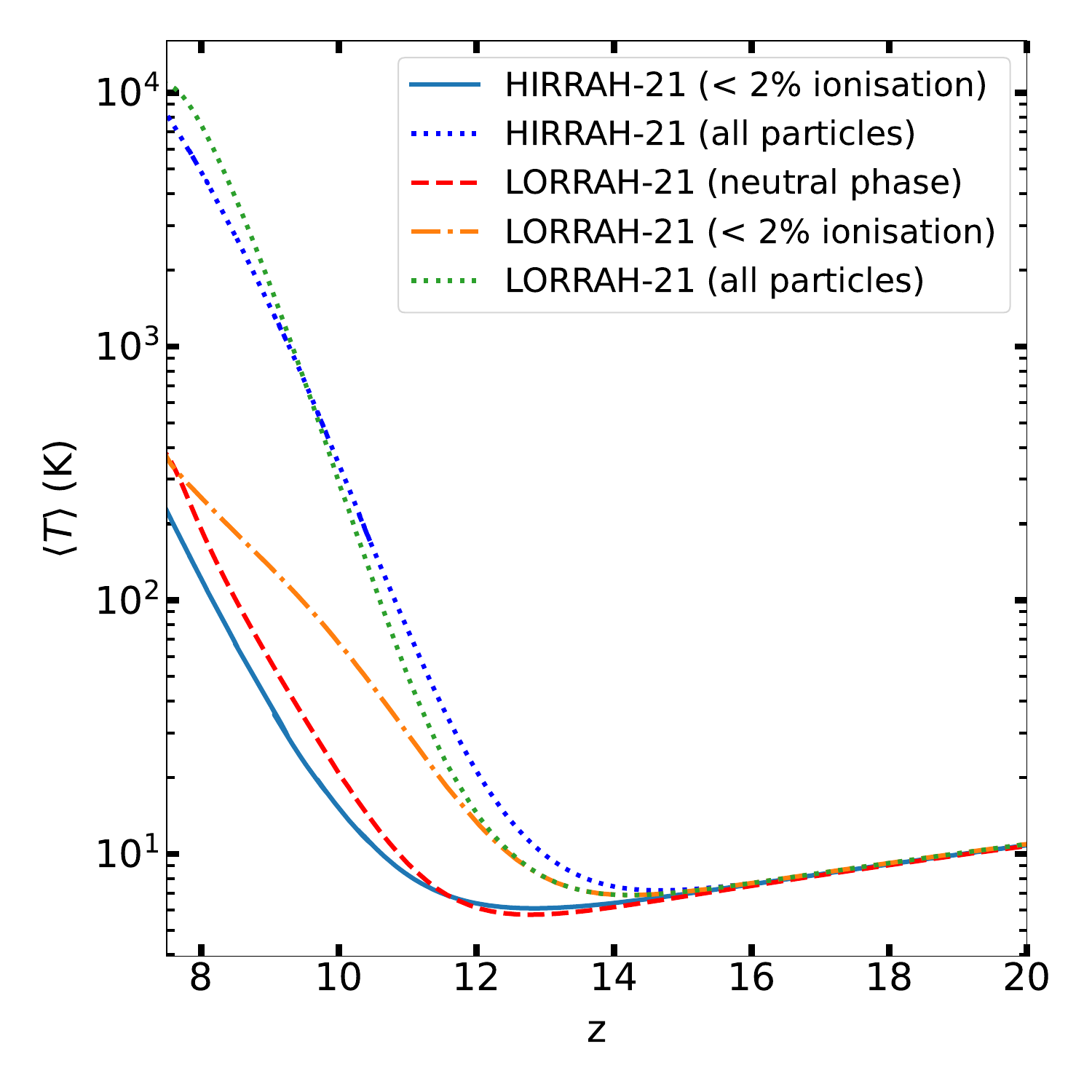}
    \caption{History of the average temperature of all particles in both LORRAH-21 and HIRRAH-21, as well as the temperature of weakly ($<2\%$) ionized particles. In the case of LORRAH-21, the temperature of the neutral phase of the particles is also shown. 
    }
    \label{fig:tp_hi}
\end{figure}

\begin{figure}
    \centering
    \includegraphics[scale = 0.32]{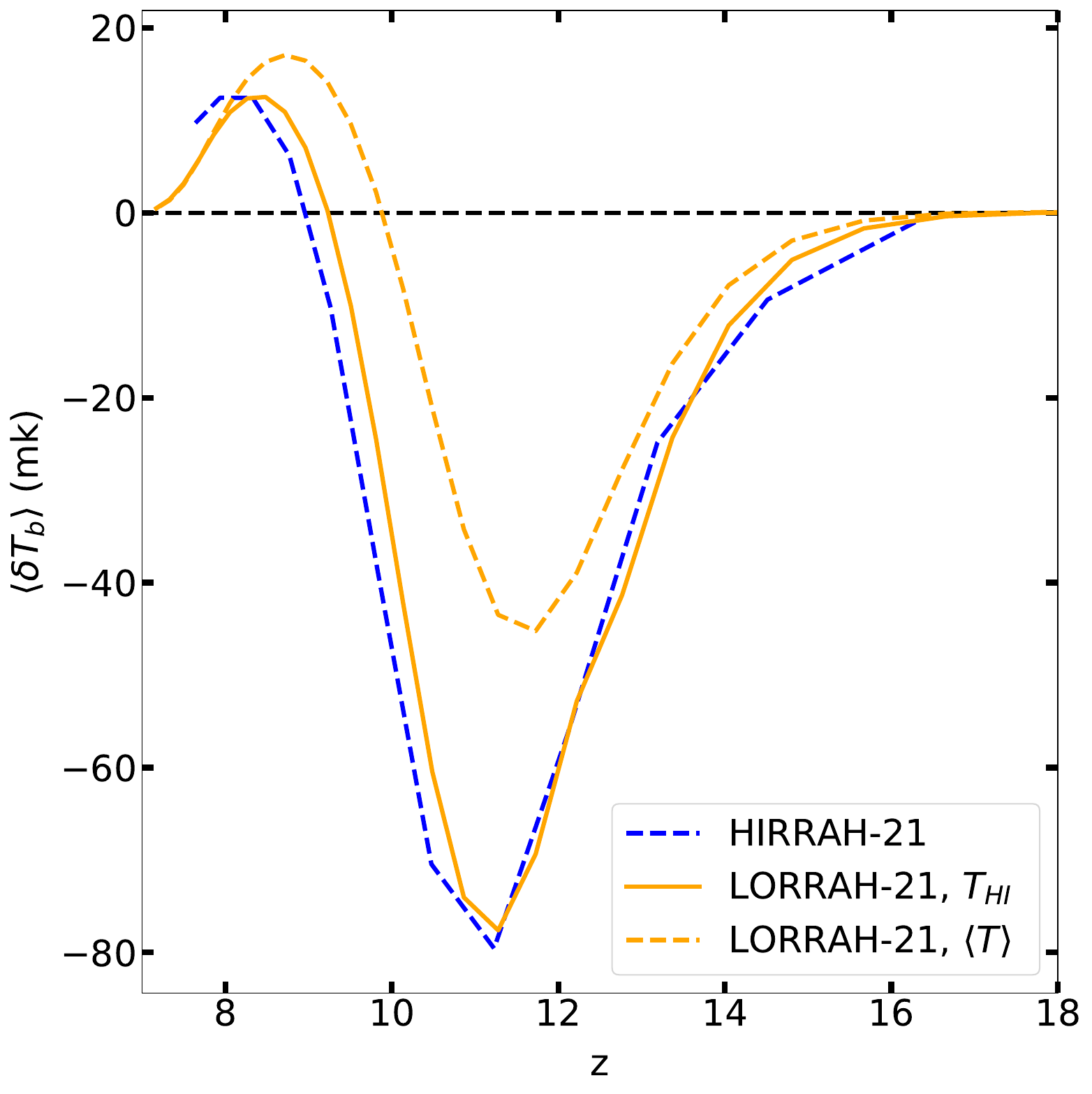}
    \caption{ Global 21 cm signal in  HIRRAH-21 and LORRAH-21, the latter calculated using either the phase-averaged temperature $\langle T \rangle$ of particles or the temperature of their neutral phase $T_{HI}$. This shows that properly taking into account the temperature of the neutral phase is critical in order to correctly model the 21 cm signal in low-resolution numerical simulations. }
    \label{fig:dtb_testpred}
\end{figure}


The other significant adaptation of the code is the computation of a specific temperature for the neutral phase within each gas particle. In previous versions of the code, only the average temperature of particles was considered (see \cite{Baek2009} for details) and used in 21 cm computations. This may have no consequence in high-resolution simulations, as ionization fronts are resolved, but becomes an issue as the resolution decreases. Low resolution implies a high number of partially ionized particles that actually represent a fully neutral phase and a fully ionized phase, and using the phase-averaged temperature then leads to a poor estimation of the intensity of the 21 cm signal. Indeed, the average temperature between the neutral and ionized gas is greater than the temperature of the neutral phase alone, often by orders of magnitude.  This does not affect codes that do not allow partial ionization, such as \textsc{21cmFAST} \citep{Mesinger2007}, and post-processing solutions have been designed to correct for this effect with the significant drawback of having to run additional simulations without X-rays \citep{Ross2016}. To compute the temperature of the neutral phase in \textsc{Licorice}, only the coupling to the dynamics, the cosmological adiabatic expansion of the gas, and the heating by X-rays were considered. This means computing the temperature evolution equation in \cite{Baek2009} with only the HI atoms:
\begin{equation}
    \frac{dT_{HI}}{dt} = \frac{2}{3 k_B n_{HI}}\left[ - \frac{3}{2}k_B T_{HI}\frac{dn_{HI}}{dt} + \Lambda \right]
,\end{equation}

\noindent where $\Lambda$ contains the X-ray heating and the adiabatic temperature evolution of the gas due to the dynamics, as well as the viscosity term of the SPH algorithm from \cite{Monaghan1992}, all computed independently for the neutral phase.  This equation describes the evolution of the internal energy of the neutral phase of a particle when interacting with the other particles in the SPH scheme.

The resulting evolution of temperatures is shown in Fig. \ref{fig:tp_hi}. These results confirm that, after the start of reionization ($z \lesssim 13$), the temperature of the neutral phase of particles in LORRAH-21 is more than an order of magnitude below their average temperature and on average significantly lower than the temperature of  weakly ionized ($\lesssim 2\%$) particles. This temperature is much closer to the temperature of weakly ionized particles of HIRRAH-21, which is used as an imperfect proxy for the temperature of the neutral phase in HIRRAH-21, which was not implemented at the time.

A decrease in resolution also affects the recombination rate $R$ of the ionized gas, as it depends on the square of the HII density. The main theoretical approach to counter this effect is to use a clumping factor. 
Different versions of the clumping factor (inspired by different works : \cite{Kaurov2014, Mao2020, Chen2020, Bianco2021}) have been implemented in \textsc{Licorice}. However, none of them managed to reconcile the reionization timings of LORRAH-21 and HIRRAH-21.

The main cause of this failure is that any error in the calibration of the parameters of the implemented clumping factor formula will cause an error in the average ionized fraction in the simulated box that increases with time. For instance, too small a clumping factor at a time step $n$ will cause too few HII to recombine, leaving too many photons per HI at time step $n+1$, further increasing the  HII density. In conclusion, none of the theoretical approaches  to the clumping factor or fit to high-resolution simulations available in the literature were accurate enough, and matching the reionization timings of our low- and high-resolution simulations required a calibration of the photon budget instead. To do so, a "step" model of the UV escape fraction $f_{esc}$ was implemented in LORRAH-21, depending on the local ionization fraction $x_{ion}$ : 

     
  \begin{align}
  f_{esc} = 
         \begin{cases}
               f_{esc, post} & \rm{if} \;  x_{ion} > x_{thresh}\\
              f_{esc, pre} & \rm{if} \; x_{ion} < x_{thresh},\\
    \end{cases}       
        \end{align}

\noindent  where $f_{esc, post} = 0.1 $, $f_{esc, pre} = 0.01 $, and $x_{thresh} = 0.01 $ are parameters of this model, calibrated on HIRRAH-21 for LORRAH-21. This is justified by the fact that the recombinations missing in LORRAH-21 occur in unresolved dense regions that are likely to surround the unresolved sources and must be ionized before letting the UV photons escape the $\sim 1 \Mpc$ region represented by a gas particle. The ionization history of LORRAH-21 using this model is consistent with that of HIRRAH-21, which is something that was not observed when the simulation code relied on a clumping factor implementation.  However, an approach using the clumping factor that leads to equivalent or more accurate results might be found in the future.

The global 21 cm signals of HIRRAH-21 and LORRAH-21  (calculated with the average temperature and with the HI temperature) are shown in Fig. \ref{fig:dtb_testpred}. As expected, the 21 cm signal is strongly modified when calculated  with the phase-averaged temperature when compared to the result obtained using the HI temperature. In addition, the 21 cm signal of LORRAH-21 calculated using the HI temperature is in good agreement with the signal in HIRRAH-21. A small mismatch occurs at $z\lesssim 11$, as heating occurs in LORRAH-21 approximately $0.15$ redshift units before HIRRAH-21, which is mainly due to the slight difference in SFRD. Over the whole redshift range, LORRAH-21 is a good approximation of HIRRAH-21, especially given the $10^4$ increase in computation speed in LORRAH-21 caused by the drop in resolution: HIRRAH-21 required $\sim 3 \times 10^6$ cpuh (which is why it was not run again with the newly implemented subgrid models) while running LORRAH-21 only takes $\sim 3 \times 10^2$ cpuh.

\section{The \textsc{LoReLi} database}\label{sec:LoReLi}

We now present the \textsc{LoReLi} database\footnote{Available at https://21ssd.obspm.fr/}, which consists of 760 simulations with $256^3$ resolution elements run in $300 \Mpc$ boxes using the code presented in the previous section, including the subgrid models for unresolved sources and the computation of the neutral phase temperature for each particle. Initial conditions, generated using the MUSIC code \citep{Hahn2011}, vary between simulations. The four-parameter space sampled by the database was designed to loosely constrain parameter values according to various probes of reionization. Here are the varied parameters and explored ranges: 

\begin{itemize}

    \item The gas-to-star conversion timescale $\tau \in [10 \, Gyr , 100 \, Gyr  ] $ and  minimum halo mass $M_{min} \in [10^8 \, \Msun, 4 \times 10^9 \, \Msun]$ are the parameters of the star formation model. 
    \item $f_{esc, post} \in \{0.05,0.1,0.2,0.3,0.4,0.5 \} $ is the escape fraction of UV radiation in particles with an ionized fraction of higher than $x_{thresh}$. 

    \item The X-ray production efficiency $f_x \in [0.1,10]$, in nine logarithmically spaced values. This drives the X-ray emissivity according to $L_x = 3.4 \times 10^{40} f_x \left(\frac{SFR}{1 \Msun yr^-1}\right) \rm{erg s^{-1} }$ \citep{Furlanetto2006a}. 

\end{itemize}

Only astrophysical parameters were varied: cosmology was kept constant across the database. Due to computation time constraints,  $f_{esc, pre}$ and $x_{thresh}$ were not varied and were kept at $0.01$, the same as in LORRAH-21. We expect these parameters to have a smaller impact on the global ionization field (except at high redshift) as the local ionization rises above the threshold early in the source life. Numerical and cosmological parameters are detailed in Table \ref{tab:numparam}.

\begin{table}[]
    \centering
    \begin{tabular}{|c|c|c|c|c|}
          \hline
          $H_0$ & $\Omega_0$ & $\Omega_b$ & $\Omega_\Lambda$ & $\sigma_8$  \\  
          \hline
          67.8 km/s/Mpc & 0.308 & 0.0484 & 0.692  & 0.815  \\
          \hline
  \end{tabular}
 \hspace{15.5cm}
  
    \begin{tabular}{|c|c|c|}
          \hline

          $L_{box} (\Mpc)$ & dt (Myr) & $N_{part}$ \\
          \hline

          294 & 7 & $256^3$\\ 
          \hline

    \end{tabular}
    \caption{Numerical and cosmological parameters of \textsc{LoReLi} simulations. }
    \label{tab:numparam} 
\end{table}

\subsection{Observational constraints }

Here we describe the regions of the parameter space that were explored and the observational constraints we used to restrict these regions. We also present the evolution of different relevant physical quantities in the \textsc{LoReLi} simulations.

\subsubsection{Star formation rate}

 \begin{figure}
     \centering
     \includegraphics[scale = 0.35]{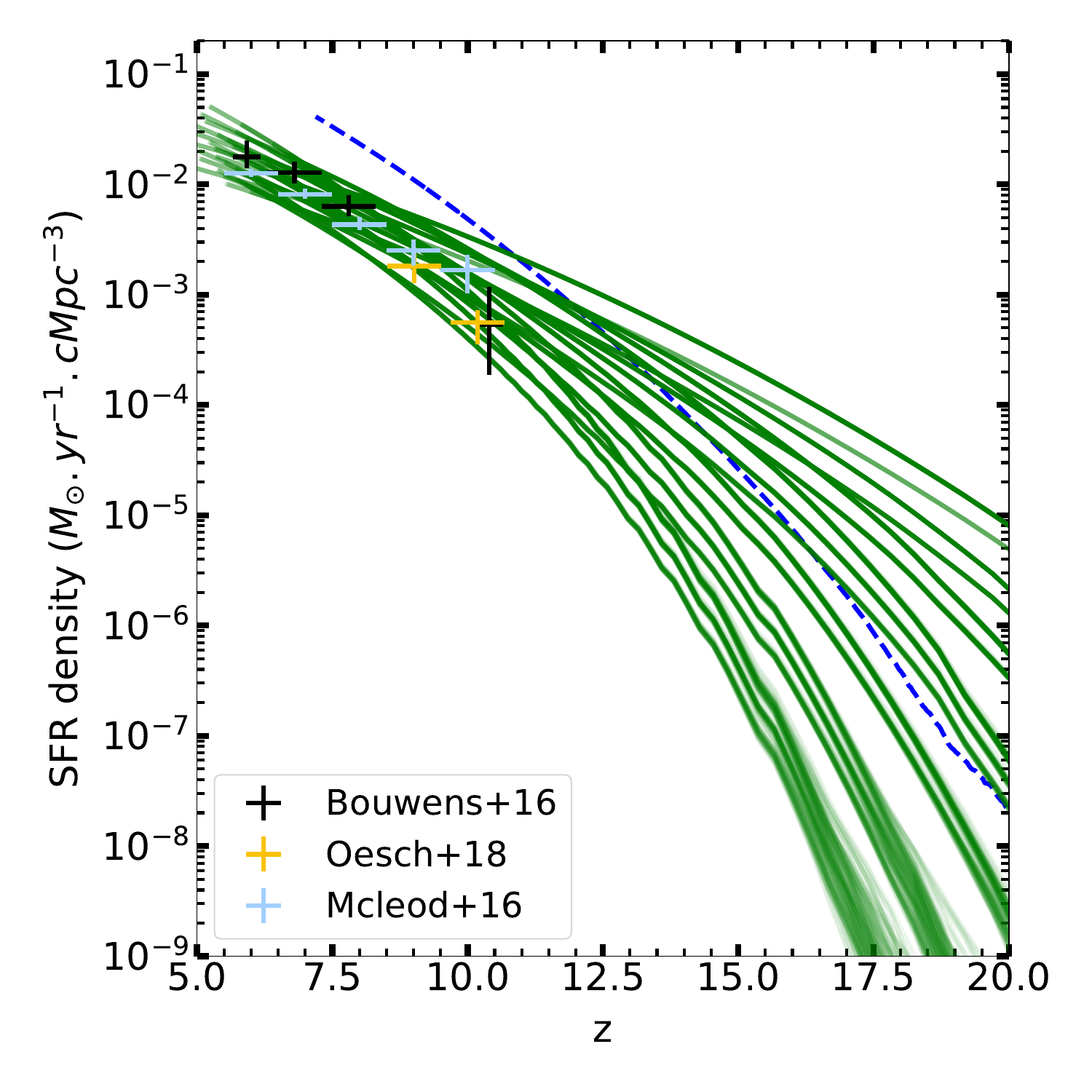}
     \caption{Star formation rate density history for the 17 ($\tau$, $M_{min}$) parameter couples used in \textsc{LoReLi} simulations (solid). The parameters were selected from a grid after a $\chi^2$ goodness-of-fit test. Only the parameter couples yielding a p-value $p$ such that $1-p < 0.95$ for at least one of the plotted observational data sets were kept from the initial grid. For reference, the SFR in HIRRAH-21 is also plotted (dashed). }
     \label{fig:sfr_loreli}
 \end{figure}

 The star formation parameters $\tau$ and $M_{min}$ were constrained using recent observations of high-redshift galaxies and SFR estimates by \cite{Bouwens2016}, \cite{McLeod2016}, and \cite{Oesch2018}. A $\chi^2$ test was performed to exclude models that fit none of the
considered observational data sets with a probability  of higher than 5 $\%$.
 Seventeen $\{ \tau, M_{min} \} $ couples were selected and the range of SFRD spanned in our database can be seen in Fig.  \ref{fig:sfr_loreli}. Measurements acquired with the JWST  are likely to give tighter and higher redshift constraints in the near future.

\subsubsection{The Thomson scattering optical depth}

  \begin{figure}
     \centering
     \includegraphics[scale = 0.35]{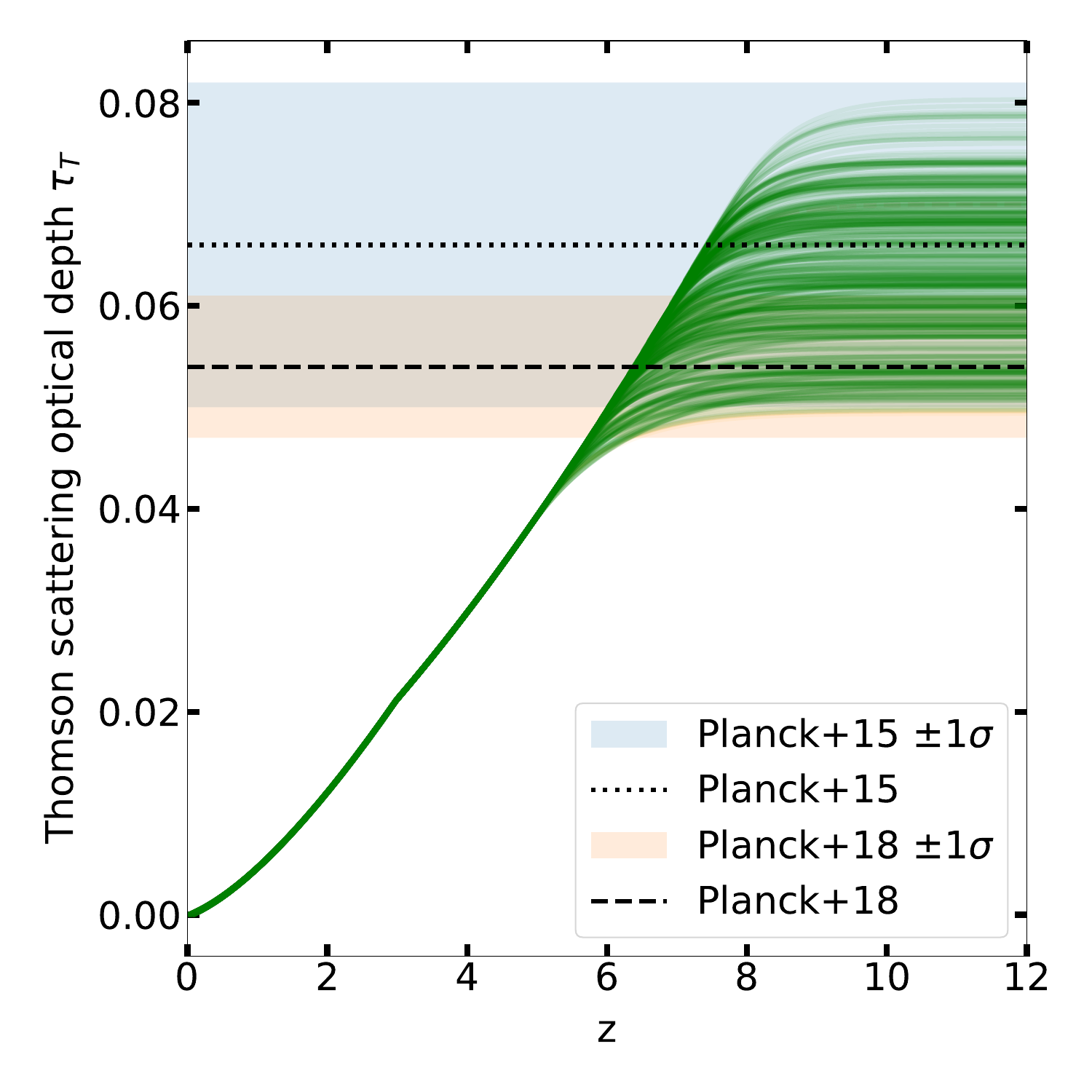}
     \caption{Thomson optical depth of the \textsc{LoReLi} signals. The ionization histories of the simulations confine the optical depths to within 1$\sigma$ of Planck's 2015 measurement (\textit{dotted}: central value, \textit{shaded region: 1$\sigma$}). As in 21SSD \citep{Semelin2017}, the second ionization of He is assumed to occur at $z = 3$.  }
     \label{fig:opticaldepth}
 \end{figure}

As late reionization scenarios are disfavored by several reionization probes \citep{Fan2006, Mitra2011,  Greig2018a}, $f_{esc}$ values were chosen so that reionization in \textsc{LoReLi} ends between $z\sim5$ and $z\sim8$.  Reionization history can also be characterized based on the Thomson optical depth $\tau_T$. The values of $\tau_T$ of the \textsc{LoReLi} simulations are shown in Fig. \ref{fig:opticaldepth} along with the results from the Planck collaboration \citep{Ade2016a,White2018}. While we did not use this observation to constrain the parameter space of \textsc{LoReLi}, nearly all models are within $3 \sigma$ of the mean value in Planck 2018 cosmology and within $1\sigma$ of that of Planck 2015. We note that the value obtained in Planck 2018 assumes simple models for reionization, and we do not exclude the couple \textsc{LoReLi} simulations that display a $\gtrsim 3 \sigma$ tension from the dataset.

  \begin{figure}
     \centering
     \includegraphics[scale = 0.35]{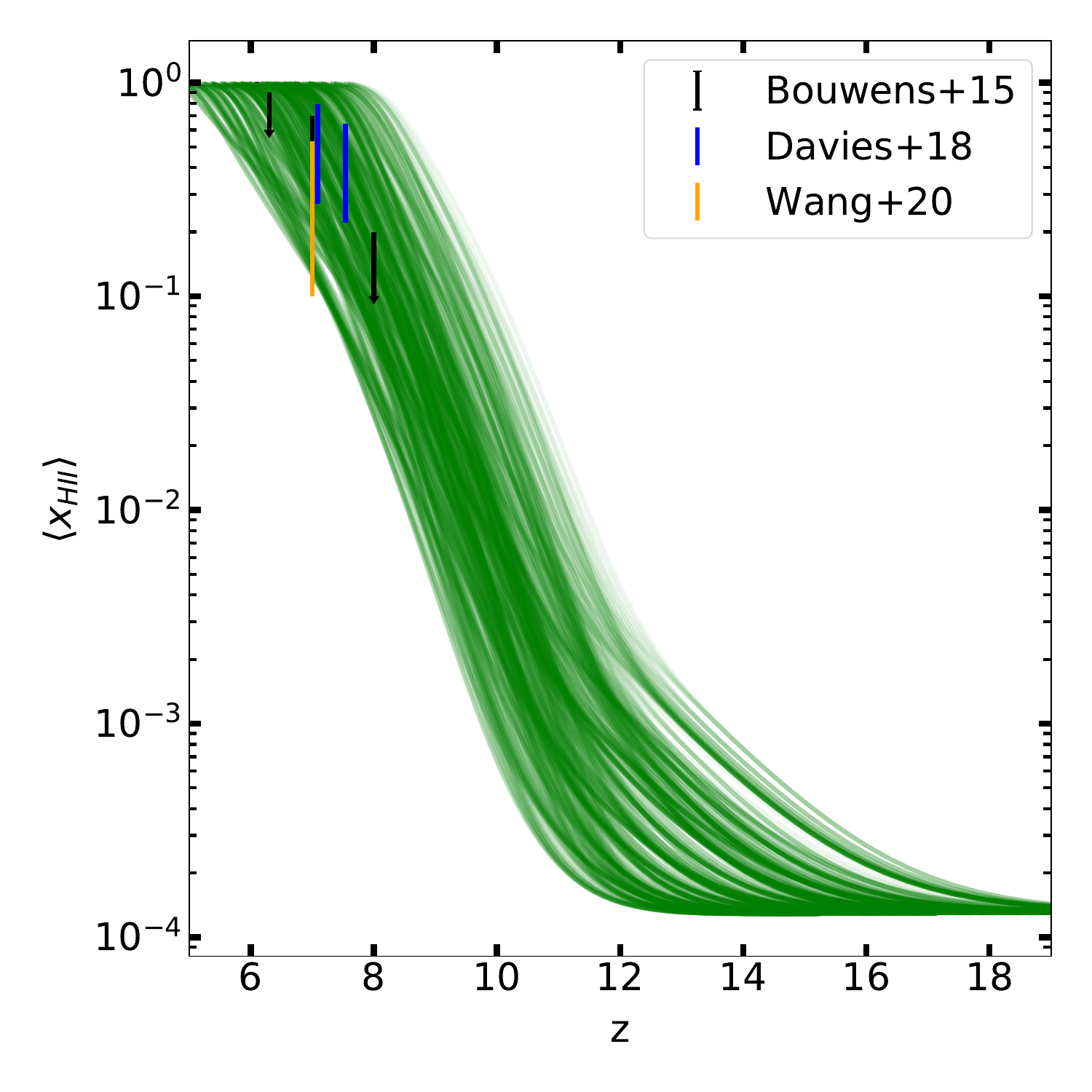}
     \caption{Average ionization fraction of the \textsc{LoReLi} models, as well as observations from \cite{Bouwens2015,Davies2018, Wang2020}. These observations were not used to calibrate the database, and reionization ends in all models between $z\sim8$ and $z\sim5$}
     \label{fig:xion}
 \end{figure}

No other constraints on reionization history were applied. In particular, no attempt was made to match observational data on the evolution of the global average ionized fraction $\langle x_{HII} \rangle$, as calibrating the escape fraction parameters for each $\{ \tau, M_{min} \} $ couple would require many simulations. For completeness, we plot the evolution of $\langle x_{HII} \rangle$ for all \textsc{LoReLi} simulations in Fig. \ref{fig:xion}.

\subsection{21 cm signals in \textsc{LoReLi}}

\begin{figure}
     \centering
     \includegraphics[scale = 0.26]{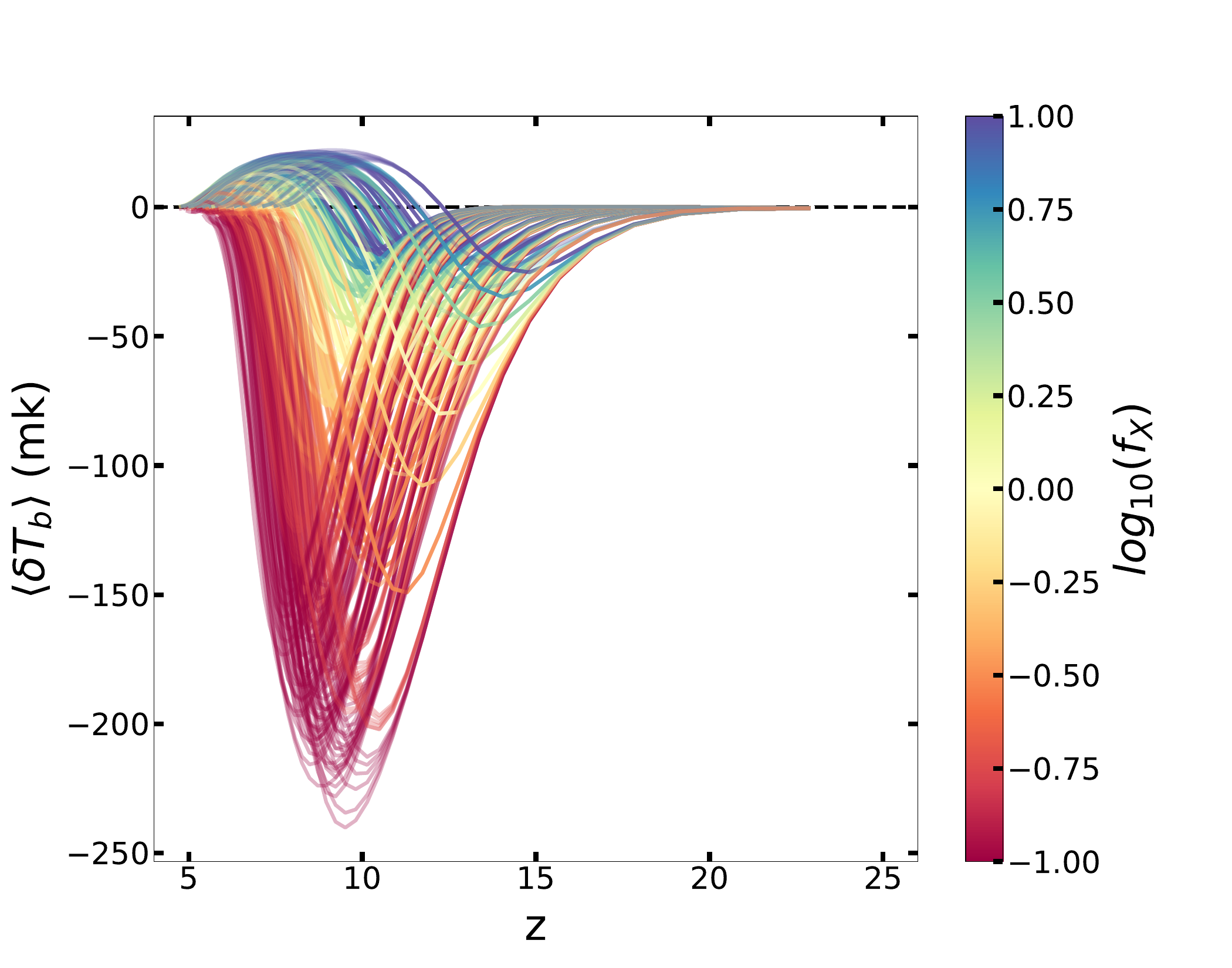}
     \caption{Global 21 cm brightness temperature of the \textsc{LoReLi} simulations. As opposed to 21SSD, global brightness temperatures are calculated from full 3D snapshots, not single slices of lightcones, and therefore do not exhibit small redshift-scale fluctuations. Color represents the value of $f_X$, the parameter which impacts the depth of the signal the most.}
     \label{fig:dtb750}
 \end{figure}

The differential brightness $\delta T_b$ of each 21 cm signal in \textsc{LoReLi} was computed according to, for example, \cite{Furlanetto2006a}:

\begin{equation}\label{eq:dtb}
\begin{split}
        \delta T_b & = 27 x_{HI}(1+\delta) \left[ \frac{T_s - T_{CMB}}{T_s} \right] \left[ 1 + \frac{dv_{||}/dr_{||}}{ H(z) }  \right]^{-1}  \\
    & \times \left[ \frac{1+z}{10}\right]^{1/2} 
    \left[ \frac{\Omega_b}{0.044} \frac{h}{0.7} \right] 
    \left[ \frac{\Omega_m}{0.27}\right]^{1/2} \, \rm{mK,}
 \end{split}
\end{equation}

\noindent  where $x_{HI}$ is the local neutral fraction, $\delta$ the local overdensity, $T_s$ the spin temperature of neutral hydrogen, $ T_{CMB}$ the CMB temperature at redshift $z$, $H(z)$ the Hubble parameter, and  $dv_{||}/dr_{||}$ the velocity gradient along the line of sight.
The spin temperature was calculated from simulated data according to the classical equation (see e.g., \cite{Furlanetto2006a}) : 

\begin{equation}\label{eq:tspin}
    T_s^{-1} = \frac{T_{CMB}^{-1} +x_cT_{kin}^{-1} +x_{\alpha}T_{kin}^{-1}  }{1 +x_c + x_{\alpha}}        
,\end{equation}

\noindent where $T_{kin}$ is the kinetic temperature of hydrogen, and $x_c$ and $x_\alpha$ the collisional and Wouthuysen-Field couplings, respectively. Collisional coupling is negligible in the considered redshift range, and therefore coupling to $T_{kin}$ is driven by the Wouthuysen-Field effect \citep{Wouthuysen1952}. $x_\alpha$ was computed on all saved snapshots for all simulations in a post-processing step using the semi-analytical code SPINTER \citep{Semelin2023}.  The luminosities of Licorice particles were arranged on a $256^3$ grid, which SPINTER takes as input. SPINTER computes spherically symmetric propagation kernels using MCMC ray tracing that accounts for scattering in the wings of Lyman-$\alpha$ line. These kernels are then convolved with the emissivity field at previous redshifts to output grids of $x_\alpha$. The outputs of SPINTER are close to those of full radiative transfer in the Lyman bands, and calculating $x_\alpha$ for a single simulation takes a few CPU hours, making the cost of this post-processing step negligible.  
We show every global 21 cm signal in Fig. \ref{fig:dtb750}. In its current state, the database contains a wide range of models that can be considered fiducial: these are shaped by conservative observational constraints and no "exotic" physics are included, such as nonstandard dark matter \citep{Barkana2018a} or extra radio background \citep{Fialkov2019}. In particular, the EDGES measurements are not compatible with any of the \textsc{LoReLi} signals, as the absorption peaks of \textsc{LoReLi} signals are shallower at lower redshifts, and have milder slopes than EDGES-compatible signals. However, some of these exotic physics, such as the extra radio background, can easily be computed in post-processing steps from \textsc{LoReLi} snapshots.

\section{Inference on power spectra}\label{sec:inference}

  \begin{figure}
     \centering
     \includegraphics[scale = 0.52]{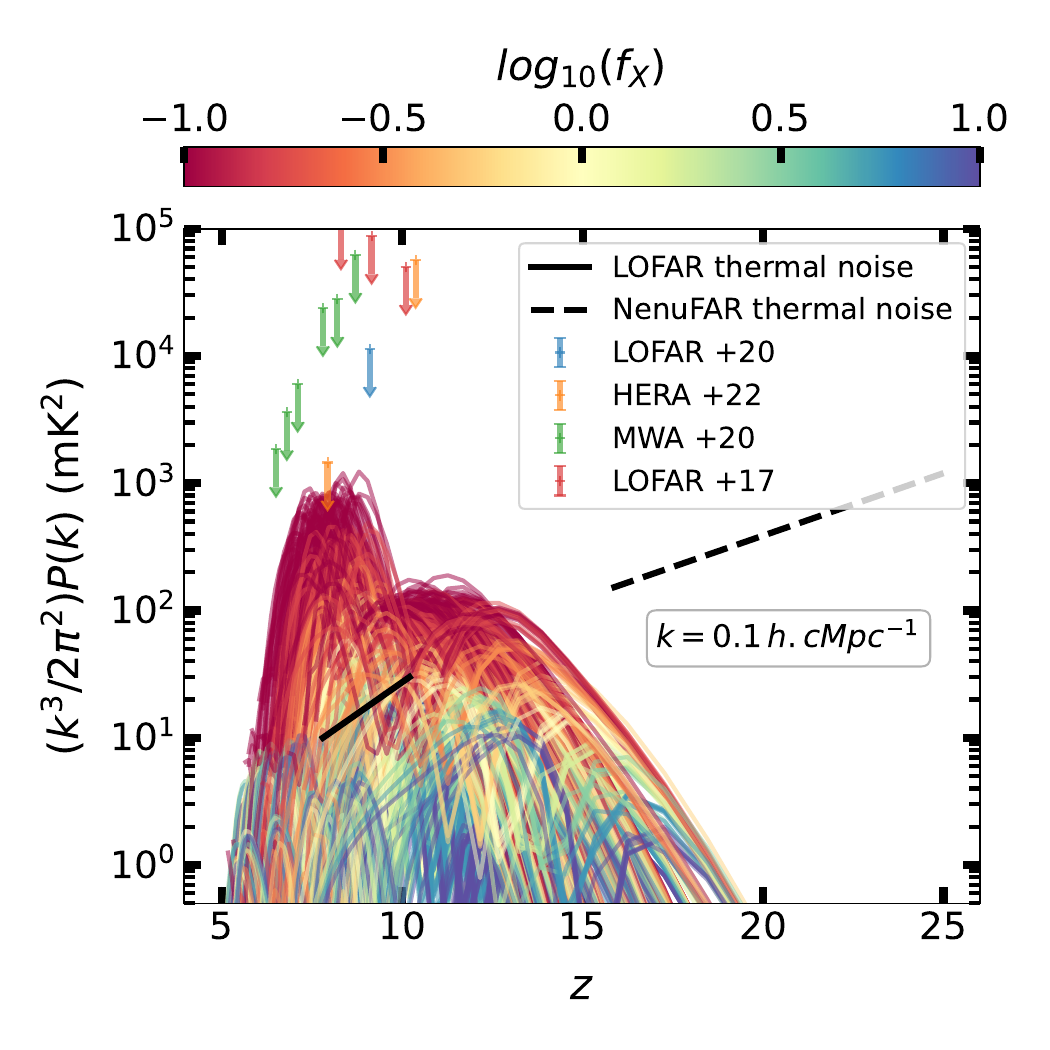}

     \caption{Power spectra of the \textsc{LoReLi} simulations at $k = 0.1 \, h/cMpc$ as a function of $z$. These are compared to recent upper limits of various instruments \citep{Patil2017, Mertens2020, Trott2020,TheHERACollaboration2022} as well as theoretical thermal noise power of 1000h observations with NenuFAR and LOFAR. 
     } 
     \label{fig:pk_vs_k}
 \end{figure}

  \begin{figure}
     \centering
     \includegraphics[scale = 0.43]{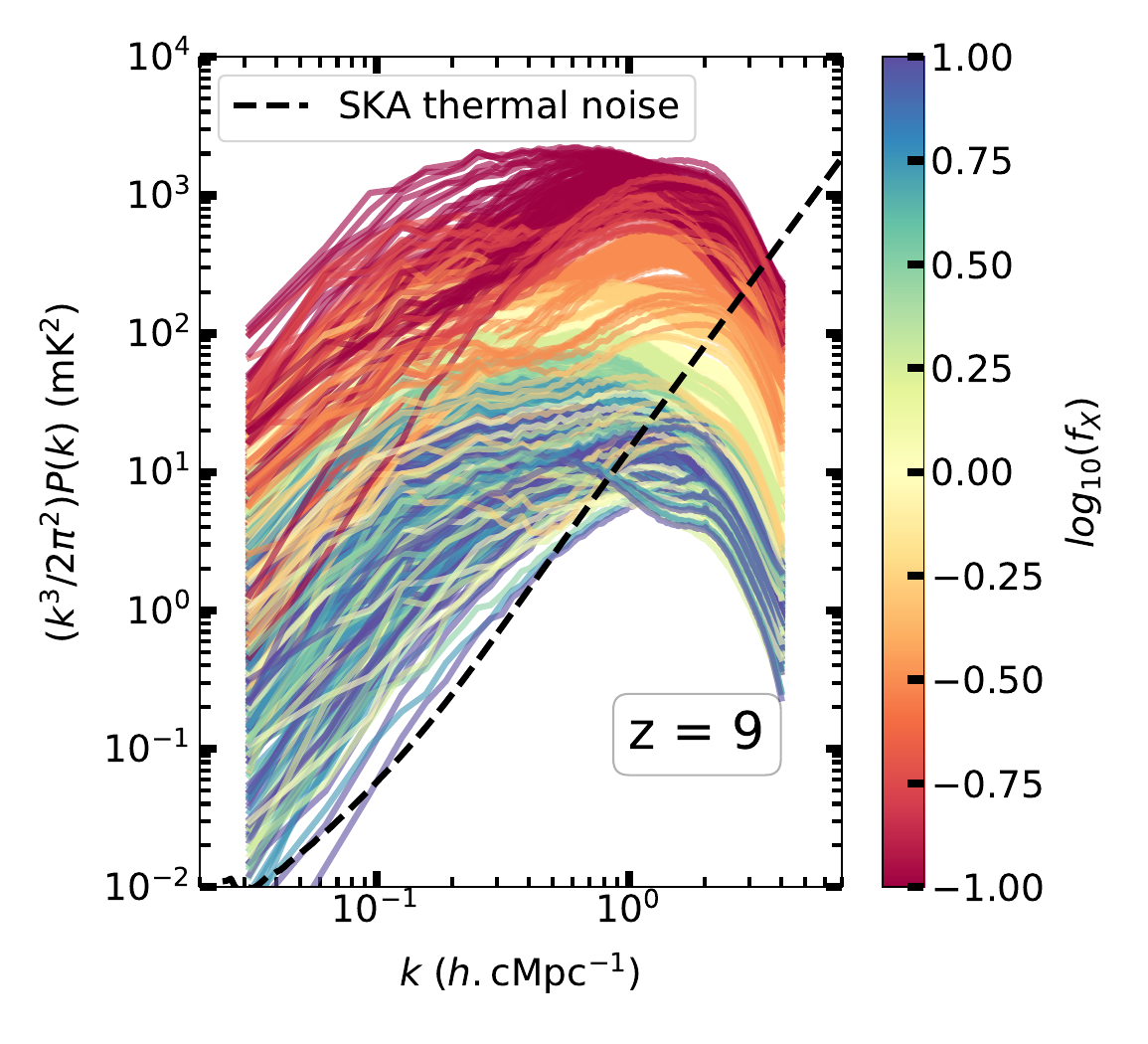}
     \includegraphics[scale = 0.43]{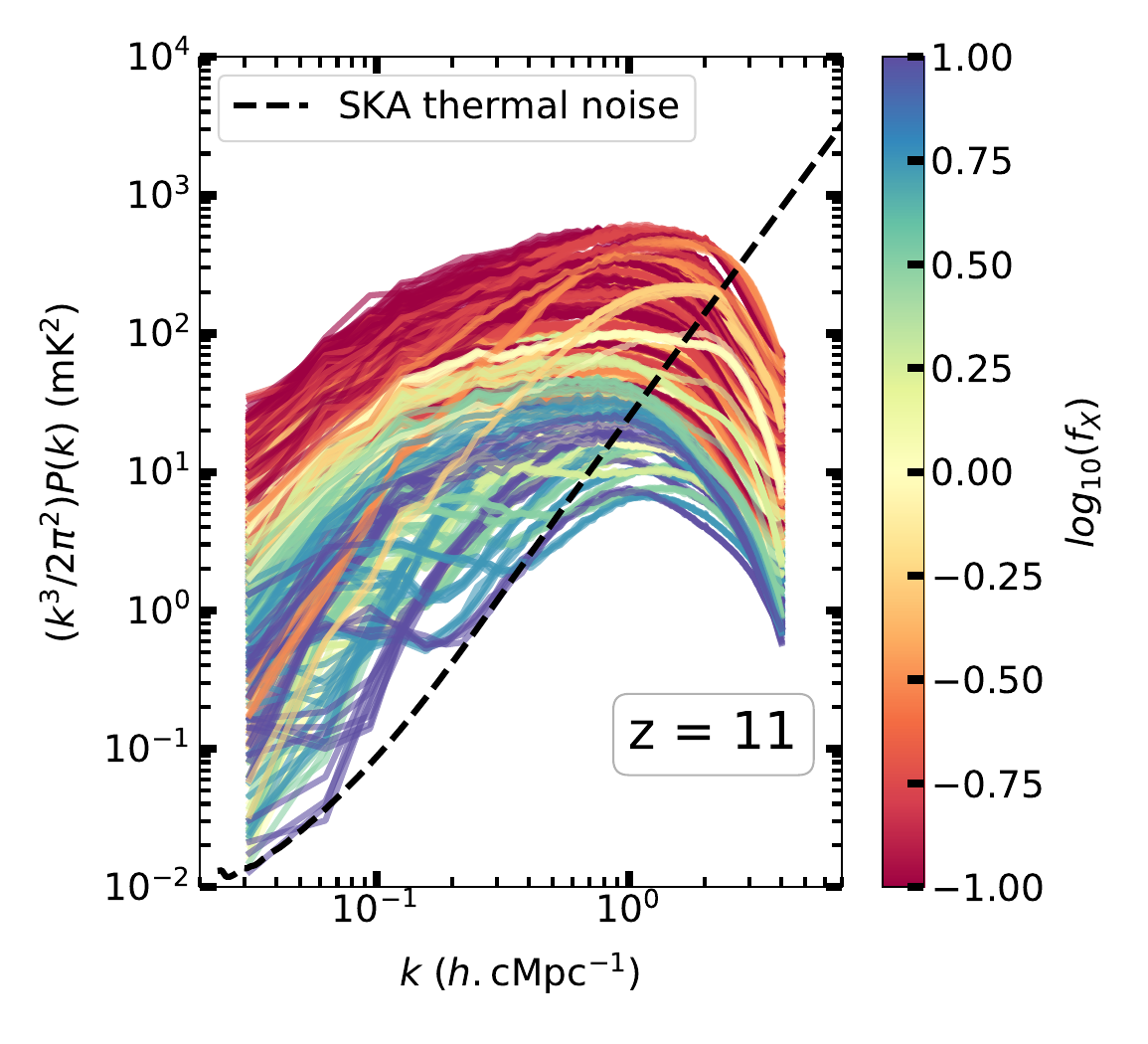}
    \includegraphics[scale=0.43]{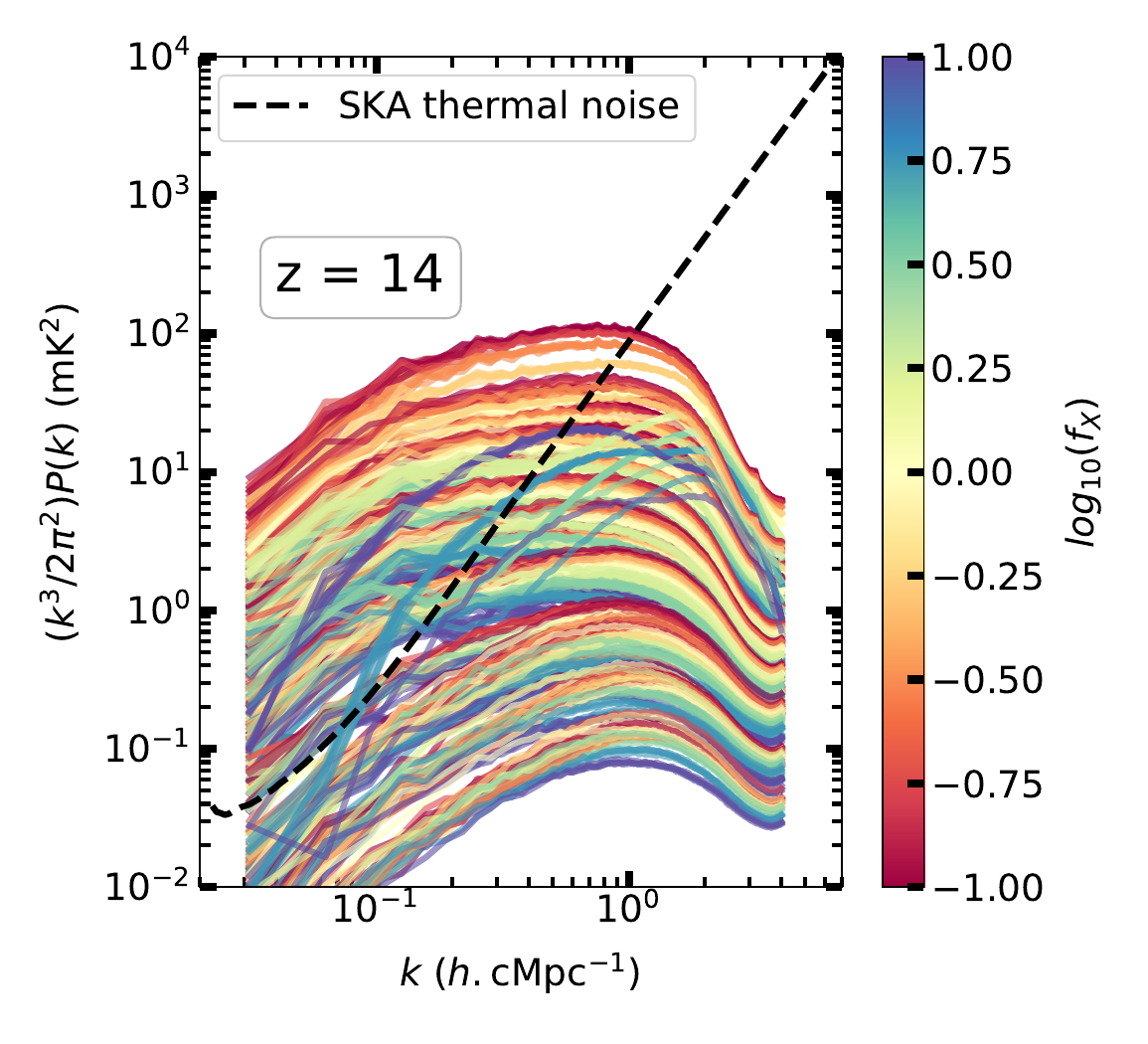}

     \caption{Power spectra of \textsc{LoReLi} models as a function of $k$ at redshifts of 9, 11, and 14, from top to bottom. The value of the SKA thermal noise (1000h of observation) is plotted to demonstrate that the \textsc{LoReLi} signals should be  detectable by the instrument.  }
     \label{fig:pk750}
 \end{figure}

In the previous section, we detail the set of simulations at our disposal. However, simulating many scenarios of the EoR is only the first step toward understanding how reionization took place in our Universe. Indeed, one must then be able to extract information from observations to determine which model is the likeliest. In this section, we present how we perform this inference step using the \textsc{LoReLi} dataset, first on mock data and then on actual observations from the HERA instrument.
\textsc{LoReLi} contains raw snapshots and lightcones containing full particle information at 55 redshifts between 53.6 and 4.97, as well as various physical quantities on 3D grids (data cubes). However, in order to compare models to future observational data, it is convenient to compress this high-dimensional data into summary statistics. The most common choice is the 3D power spectrum of the 21 cm signal, which various instruments are currently trying to measure \citep{Patil2017, Mertens2020, Trott2020,TheHERACollaboration2022}. While it does not contain non-Gaussian information, and summaries representing complementary information do exist, we  focus on the power spectrum in the following. We show the power spectra of the \textsc{LoReLi} simulations in Figs. \ref{fig:pk_vs_k} and \ref{fig:pk750}, as well as upper limits from recent observations, and expected thermal noises of various instruments. The power spectra were calculated using the coeval cubes of the signal and not the lightcones. This approximation can have a significant (up to $\sim50\%$) effect on the spectrum \citep{Datta2012, Datta2014} and taking this into account will be at the center of future improvements of our method.
        As expected from classical 21cm theory, at higher redshift, the amplitude of the spectra is mostly driven by the Lyman-$\alpha$ coupling, and becomes primarily dictated by the value of $f_X$ at lower redshifts.
        We anticipate that improvements of these upper limits in the coming years will soon allow tight constraints on our models.


\begin{figure*}
         \centering
     \includegraphics[scale = 0.85]{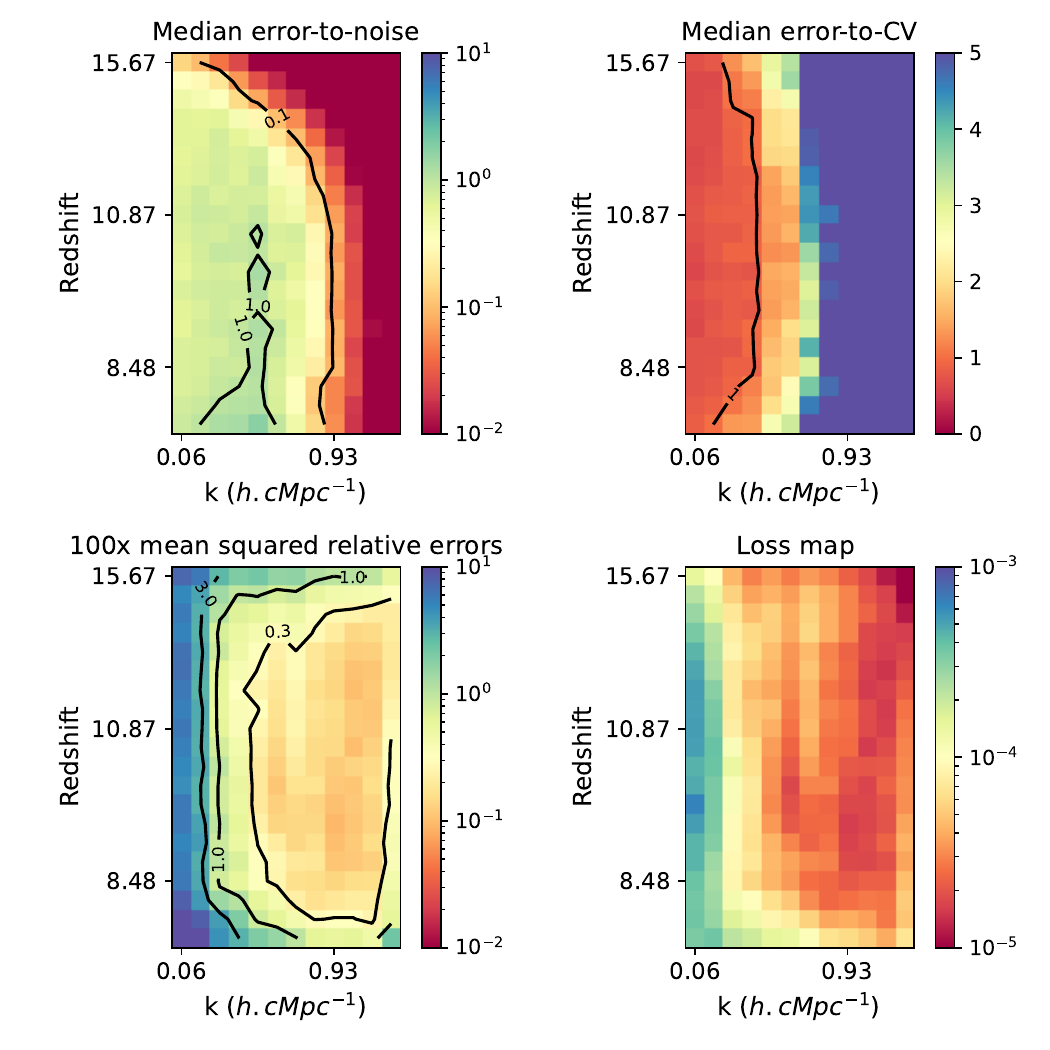}
     \caption{Diagnostics of the performance of the emulator. \textit{Top left}: At each $k,z$, median across the dataset of the emulator error-to-thermal-noise ratio. \textit{Top right}: At each $k,z$, median across the dataset of the emulator error-to-cosmic-variance ratio. The error increases with cosmic variance, as the deterministic network cannot replicate the noise induced by cosmic variance. \textit{Bottom left: } 100x mean squared relative error across the dataset between the emulator prediction and the training sample. \textit{Bottom right: } Loss at each k, z. Faint signal at high z causes low loss but high relative error. Cosmic variance causes high loss and relative error at low k. Additionally, we checked that the  mean square relative error  exhibits no systematic trend depending on the values of the astrophysical parameters.  }
     \label{fig:emul}
\end{figure*}

\begin{figure*}
    \centering
    \includegraphics[scale = 0.8]{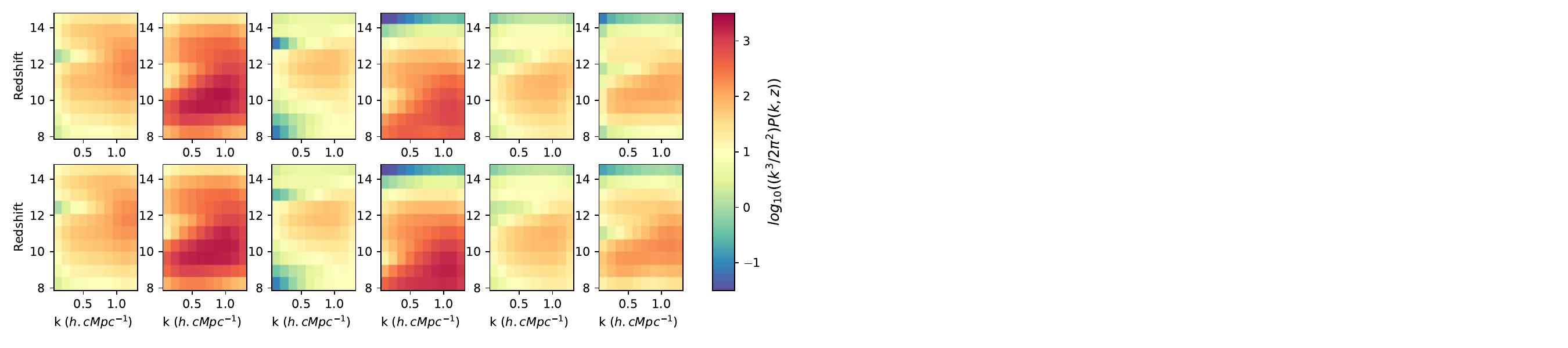}
    \caption{Randomly selected examples of true (top) and emulated (bottom) \textsc{LoReLi} signals qualitatively showing the resemblance between emulated and simulated spectra.}.  
    \label{fig:mozaic}
\end{figure*}


\begin{figure*}
         \centering
     \includegraphics[scale = 0.73]{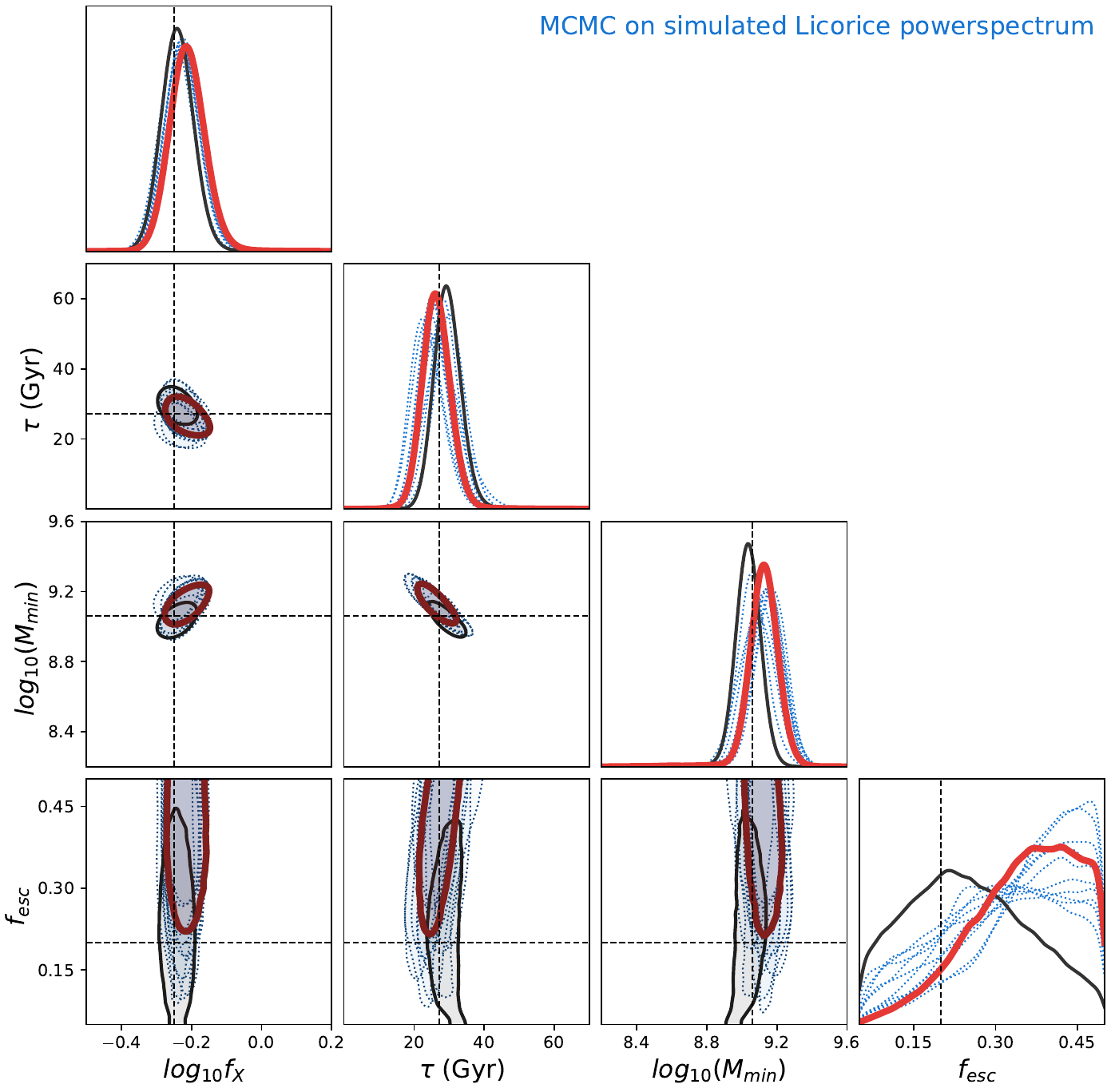}
     \caption{Results of multiple inferences done using several versions of \textsc{LorEMU}, trained with different weight initializations (dashed). On the 2D panels, only the 1$\sigma$ contours are shown for clarity. The blue dotted lines are the posteriors obtained using each emulator. The thick red line shows the posterior obtained using the average emulator on the noised signal, while the inference on the noiseless signal is shown in black.   }
     \label{fig:inference_simu}
\end{figure*}

\begin{figure*}
         \centering
     \includegraphics[scale = 0.78]{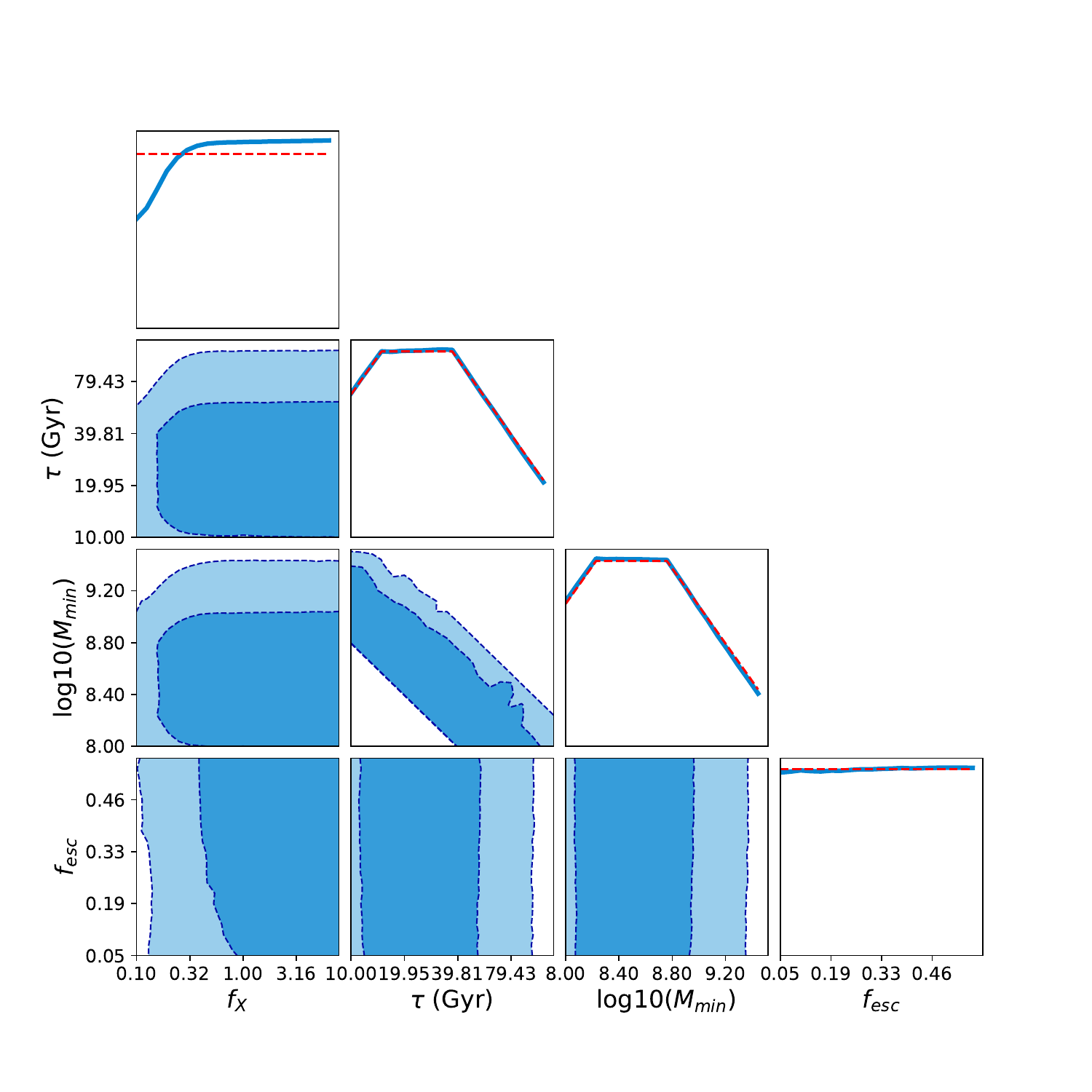}
     \caption{Results of inference on HERA data using \textsc{LorEMU}. On the 1D panels, the priors are in red and the posteriors in blue. Contours represent $68\%$ and $95\%$ levels of confidence.}
     \label{fig:inference_hera}
\end{figure*}

Classical Bayesian inference, for example through the use of MCMC, requires large numbers of forward modeling instances ($\gtrsim 10^5$), often sequentially. Given the computation cost of a single \textsc{Licorice} simulation, this is unreasonable even at $256^3$ resolution. 
In order to perform parameter inference on this power spectra dataset, our method consists in using \textsc{LoReLi} as a training sample for a neural network that will function as an emulator of \textsc{Licorice}, and then performing classical MCMC inference using the emulator as the model. Indeed, once the emulator is trained, the computation cost of producing a single signal drops to a few milliseconds, which allows a sufficiently large number of steps to be completed in the Markov chain in a few hours of runtime.

\subsection{Data preprocessing}

\subsubsection{Data properties}

In order to efficiently train the network, it is necessary to preprocess the data. The power spectra were computed for a list of 32 redshifts, with power spectra being set to zero for $z$ between the redshift of full reionization  and the last redshift bin. However, in order to reduce the dimensionality of the problem, we restrict our analysis to $k = [0.23, 0.33, 0.46, 0.66, 0.93, 1.31, 1.86, 2.64 \rm{h/cMpc}]$ and $z = [15.67, 14.05, 12.77, 11.73, 10.87, 10.14, 9.
51,8.96]$. The $k$ bins width is defined such that the bins do not overlap. The 21 cm power spectra span many orders of magnitude, which tends to hinder the training of networks. Therefore, the logarithm of the power spectra was used before normalizing by dividing by the maximum value across the dataset.

\subsubsection{Noise in the signals}

For an inference framework to have any practical application, it must be performed on signals affected by instrumental effects. Here we neglect foreground residuals and other systematics and focus on the SKA thermal noise for an observation time of 100h. As in \cite{Doussot2019}, and following \cite{Mcquinn}, the standard deviation of the power spectrum caused by SKA thermal noise was computed as 

\begin{equation}
    \delta P(k,z) = \left[\sum_{|k| = k}\left(  \frac{1}{ \frac{Ax^2 y }{\lambda(z)^2 B^2} C(\textbf{k},\textbf{k})   } \right)^2 \right]^{-1/2}
,\end{equation}

\noindent  where $\lambda$ is the observed wavelength, $A$ the area of a 256-antenna station, $B = 10 \, \rm{MHz}$ is the bandwidth, $x$ the comoving distance to the observed redshift, $y$ the depth of field, and $C$ the detector covariance matrix. In order to noise a signal, we simply add a realization of a Gaussian random variable of mean 0 and standard deviation $\delta P(k,z) $ for each $k,z.$

This noise is added to another form of stochasticity, the cosmic variance of the simulation box (CV), which is implicit to our setup. We recall the equation found in \cite{Mcquinn} to compute the covariance matrix of CV, which is added to the detector noise covariance matrix in the previous equation: 
\begin{equation}
     C_{CV}(k_i, k_j) \approx     P(k_i)\frac{\lambda^2 B^2}{Ax^2 y }\delta_{ij}
.\end{equation}


Marginalizing over CV for each point in parameter space  requires hundreds of simulations at each point and is not feasible using \textsc{LoReLi}. However, as CV mainly affects the largest scales, its effect can be at least partially avoided by focusing the analysis on $k\gtrsim 0.1 \,\rm{h/cMpc}$. This motivates us to focus our analysis on a region in $k$-space more affected by thermal noise, which is far less computationally expensive to generate.
It is also likely that a machine learning method that tends to interpolate between the \textsc{LoReLi} simulations, such as the one presented in the following section, smooths over the effects of CV altogether.

\subsection{ \textsc{LorEMU}}

\begin{table*}[]
    \centering
    \begin{tabular}{|c|c|c|}
        \hline 
         \textbf{Layer type} & \textbf{Activation function} & \textbf{Regularization}  \\
         \hline             

         Dense , 512 neurons & Leaky ReLu ($\alpha = 0.05$)& L1 ($\lambda = 5\times 10^{-6}$)  \\
        Dense, 512 neurons & Leaky ReLu ($\alpha = 0.05$) & L1 ($\lambda = 5\times 10^{-8}$)  \\
        Dense, 512 neurons & Leaky ReLu ($\alpha = 0.05$)& None  \\
        Dense, 1 neuron & Sigmoid & None  \\
        
    \hline\hline                 

    \hline
    \textbf{Optimizer} &  \textbf{Loss function} & \textbf{Batch size} \\
    \hline
    Adam (learning rate = $5\times10^{-4})$ & Mean Squared Error & 32\\
    
    \hline 
    \end{tabular}
    
    \caption{Architecture and hyperparameters of \textsc{LorEMU}.   }
    \label{tab:archi}
\end{table*}

\textsc{LorEMU}, an emulator of the \textsc{Licorice} power spectra, was trained using the \textsc{LoReLi} simulations. The network is a multilayer perceptron that takes the four astrophysical parameters and a (k, z) pair as inputs, and outputs a noiseless power spectrum value. The full architecture and parameters are shown in Table\footnote{The detailed signification of the technical terms can be found at https://keras.io/} \ref{tab:archi} and were implemented using the Keras framework \citep{Chollet2015}. LorEMU was trained in a supervised manner to predict the value of the (noiseless) power spectrum given the inputs $k,z$, and the astrophysical parameter values. Training took place over 200 epochs, using a batch size of 32, and minimizing a mean squared error (MSE) loss function using the Adam optimizer with a learning rate of $5\times10^{-4}$. The training set was composed of a randomly selected $75\%$ of the 760 spectra, while the remaining  $25\%$ of the spectra were set aside to constitute the test set. 

This network is deterministic. In particular, this means the output power spectra do not include a proper contribution by CV. It is hypothesized that the emulator learns to average over the CV affecting the large scales, as the spectra in the training set that are close in parameter space have been generated using different initial conditions of the density field and therefore have different CV realizations. Various quantities describing the accuracy of the emulator are  depicted in Fig. \ref{fig:emul}. As indicated by the bottom left plot, the emulator performs especially well at high $k$ and low redshifts. At large scales, the error is caused by the fact that the emulator cannot predict CV, whereas spectra in the test set include a CV realization (see top right plot). At high redshift, the signal is faint, resulting in low loss but high relative error (see bottom plots). The rise in relative error at the very lowest redshift may partially be due to simulations approaching the end of reionization, but is in this case accompanied by a rise in the loss function, which is more difficult to explain. A similar issue was found to occur by  \cite{Jennings2019}, and this region in (k,z) may be the focus of future improvements of the emulator.  Overall, our emulator outperforms that of \cite{Jennings2019} by a factor of $\sim 2$.
We show examples of simulated and emulated signals in Fig. \ref{fig:mozaic}.

\subsubsection{Inference on mock data}

\textsc{LorEMU} allows the generation of a $8\times8$ power spectrum in milliseconds and is therefore a suitable method for forward modeling in an MCMC inference pipeline. 
Figure 14 shows examples of Bayesian inference on a simulated spectrum \footnote{ calculated using the following astrophysical parameters: $f_{X}=0.56$, $\tau=27 \,Gyr$, $M_{min} = 10^{9.06}  \Msun$, $f_{esc,post} = 0.2$  }, assuming perfect foreground removal and noise corresponding to a 100h SKA observation, as explained previously. The inferences were performed using the \textsc{emcee} Python package \citep{Foreman-Mackey2013}. The log-likelihood of a model $y$ with astrophysical parameter set $\theta$, with respect to data $x$, and with total variance $\sigma_{tot}$, can be explicitly written: 
        
\begin{equation}
    logL(y | \theta) = - \sum_{k,z} \frac{1}{2} \left( \frac{x(k,z) - y(\theta, k ,z )}{\sigma_{tot}(k,z)} \right)^2
.\end{equation}
It is worth noting that this form of the likelihood assumes that Fourier modes are independent variables, an assumption discussed in \cite{Prelogovic2023}  for example.  While these latter authors show that using a diagonal covariance for the cosmic variance can lead to biased inference results, they also show that the nondiagonal terms are mostly erased by the level of their thermal noise, which is an order of magnitude below that used in our work. Therefore, we expect no significant contribution from these nondiagonal terms in our case. In our case, $\sigma_{tot}$ is 
\begin{equation}
    \sigma_{tot} = \sqrt{\sigma_{thermal}^2 + \sigma_{CV}^2 + \sigma_{train}^2},
\end{equation}
and includes, at each bin of $k,z,$ the cosmic variance $\sigma_{CV}$, the variance of the SKA thermal noise $\sigma_{thermal}$, and the "training variance" $\sigma_{train}$, which represents the variance in the training of the network and is detailed below.

According to Bayes theorem, the posterior distribution of the parameters can be written:

\begin{equation}
    p(\theta | y) \propto L(y | \theta)\pi(\theta)
,\end{equation}

\noindent  where $\pi(\theta)$ is the prior over the astrophysical parameters.

In order to estimate the error induced by an imperfect training of the emulator,  nine different versions of \textsc{LorEMU} were trained. Each version uses the same architecture, and was trained using different weight initialization and with  different random splits of the data to constitute the test and training sets.  The rest of the training pipeline was kept unchanged. Each one of these trained emulators was subsequently used as the forward simulator in the MCMC pipeline, producing the different posteriors of Fig. \ref{fig:inference_simu}. The inferences were done using 160 walkers, randomly initialized in the prior, and approximately $ 800 000$ total steps. 
The prior on the astrophysical parameters is flat within the region in parameter space explored in \textsc{LoReLi} and zero outside this region.

The inferences were all performed on the same simulated signal noised with the same 100h SKA noise realisation, and therefore the differences between inferences are due to the different weight initializations and training stochasticity between the different versions of \textsc{LorEMU}.
This training variance can be defined at each $k,z$, and astrophysical parameters set $\theta$ as 

\begin{equation}
  \sigma_{train}^2 = \frac{1}{N}  \sum_N \left( P_{predicted}(k,z,\theta) - P_{target}(k,z,\theta) \right)^2  
,\end{equation}

where $P_{target}$ and $P_{predicted}$ are the training spectra and the outputs of the emulator, respectively, and the sum is taken over a large number $N$ of emulators trained with different weight initializations. Rigorously evaluating this variance would not only require a large number of emulators, but also a target signal for any value of $\theta$, while we have signals only for the parameter values present in the LoReLi database. To address those difficulties, we chose to average this quantity over $\theta$. The result is an approximation of the training variance. However, this is justified by the fact that the error shows no clear dependency on parameters. This variance adds a bias in the inference with single emulators, and putting the training variance in the likelihood widens the confidence contours. While negligible for $f_X$, the effect of this training variance is comparable to that of the SKA 100h thermal noise for the SFR parameters. In an attempt to marginalize over the random initializations, Fig. \ref{fig:inference_simu} shows a posterior obtained using the average predictions of the nine emulators in the MCMC framework, following the bagging method commonly used in machine learning. In order to demonstrate that, for the average emulator, the biases of the predictions are negligible compared to the uncertainty induced by the noise,  Fig. \ref{fig:inference_simu} also shows the posterior applied to the noiseless signal (but with the SKA 100h thermal noise variance still included in the likelihood). We see that, in this case, the confidence contours are well centered on the target parameter values. This also indicates that our inference results are only weakly affected by our choice of using a diagonal covariance in the likelihood.

$\tau$ and $M_{min}$ are unsurprisingly strongly anticorrelated, as their effects on star formation are degenerate. Similarly, $f_X$ is correlated with $M_{min}$ and anticorrelated with $\tau$, as X-ray emissivity is proportional to SFR.  We note that these effects are unlikely to stem from our choice of priors, because they are visible in confidence contours far away from the flat prior boundaries. Additionally, $f_{esc}$ is the least strongly constrained parameter, with large contours and non-negligible likelihood over the whole prior. Except for $f_{esc}$, the true values of the astrophysical parameters are within the $1\sigma$ contours obtained after inference with the average emulator.

\subsubsection{Inference on recent HERA data}

In the previous section, we validated our method by applying it to mock observations generated from our simulated data. We now turn to real data, and provide an analysis of the most recent HERA observations \citep{TheHERACollaboration2022}, at $z= 10.4$ and $z = 7.9$. As indicated in Fig. \ref{fig:pk750}, the most recent HERA upper limits are currently the only ones capable of constraining our data set.
We follow the procedure detailed in \cite{Abdurashidova2022}  (thereafter HERA2022), which we briefly summarize here. The observed data are assumed to be the 21 cm signal, to which systematic uncertainties (typically foreground residuals and radio frequency interference) are added. The power
of the systematic uncertainties is supposed to be positive, and marginalizing over them yields the likelihood:

\begin{equation}
    L(y | \theta) = \prod_{k,z} \frac{1}{2} \left[ 1 + \rm{erf} \left( \frac{x(k,z) - y(\theta, k ,z )}{\sqrt{2}\sigma(k,z)}\right) \right].
\end{equation}

The emulator was not retrained on spectra calculated with the $k$ bins of HERA data with its window function. The posterior distribution of the astrophysical parameters obtained through MCMC inference is shown in Fig. \ref{fig:inference_hera}. The only parameter for which the posterior distribution is different from its prior is $f_X$. Qualitatively, our conclusions match those of HERA2022: a cold reionization of the Universe is an unlikely scenario. 
However, quantitatively,  the constraints we put on $f_X$ appear looser than in HERA2022.   In figure 6 of HERA2022, which shows the inference results as obtained using \textsc{21cmFAST}, the likelihood decreases for $f_x < 1$, while we only observe such a decrease for $f_X < 0.5$. This may stem from a difference in the prior on the SFR parameters. Indeed, the observational constraints on SFR applied to \textsc{LoReLi} and \textsc{21cmFAST} only exist at $z\leq10$, leaving the functional form of the SFR evolution completely free at higher redshift. Different posteriors can result from this, as seminumerical codes typically assume that stars are formed instantaneously as some fraction of the collapsed gas. As detailed in Sect. \ref{sec:Licoricecode}, in \textsc{Licorice}, the stellar fraction evolves with time. If the stellar fraction in \textsc{Licorice} is typically lower than in seminumerical codes at high redshift and higher at lower redshift in order to be consistent with the same observational constraints on  SFRD, then this difference in the inference result is expected.
Unfortunately, as the codes function in significantly different ways with different parametrizations, understanding the difference in the predicted posteriors of both inferences would require an in-depth comparison of \textsc{Licorice} and \textsc{21cmFAST} (as well as with other codes) that does not yet exist and lies beyond the scope of this work. 
Regarding figure 14 of HERA2022, which was obtained using the seminumerical code described in HERA2022, an additional explanation may be that the authors marginalize over $f_r$, a parameter that controls the intensity of an exotic radio background, while we implicitly fix $f_r$ to zero as the only background we consider is the CMB.
 
As a final note, the inference was also attempted on upper limits from LOFAR \citep{Mertens2020} using the exact same methodology. As could be expected from Fig. \ref{fig:pk750}, no difference between priors and posteriors could be established in that case. 
\section{Conclusions}

In this work, we present new functionalities of the \textsc{Licorice} simulation code that allow low-resolution simulations of the EoR that reproduce many features of high-resolution simulations. These new functionalities are twofold. One is an implementation of the conditional mass function formalism as a way to statistically estimate the mass of unresolved dark matter halos. We then include the gas contained in these halos to compute star formation in low-resolution simulations, which would otherwise be severely underestimated. Using the Sheth-Tormen formulation of the CMF, we find excellent agreement with $1024^3$ and $2048^3$ simulations.

The other modification is a calculation of the temperature of the neutral phase of each particle. Due to poorly resolved ionization fronts in low-resolution simulations, we find that using the phase-averaged temperature of the particles leads to a significant overestimate of the intensity of the 21cm signal. Using the neutral phase temperature  leads to good agreement between results at low and high resolution.

Together, these improvements allow physically reasonable Licorice simulations at $256^3$ resolution that run in approximately $300$ cpuh. Consequently, this makes running many Licorice simulations computationally feasible. Therefore, we present \textsc{LoReLi}, a growing dataset  of 760 \textsc{Licorice} simulations with $256^3$ particles in $300 \Mpc$ boxes. Full particle data are saved at 32 redshift values, and \textsc{LoReLi} spans a four-parameter space: the escape fraction $f_{esc, post}$, the X-ray emissivity $f_X$, the minimum mass of star forming halos $M_{min}$, and the gas-to-star conversion timescale $\tau$. The latter two parameters were calibrated using constraints on the SFRD, and the first was chosen so that reionization ends in all models at between $z\sim5$ and $z\sim8$.  \textsc{LoReLi} therefore contains a variety of standard models of the EoR.

In the first application of this dataset, we summarize our data into independent power spectra values at 8 $k$ and 8 $z$ values. We then present \textsc{LorEMU}, a neural network trained on \textsc{LoReLi} to produce accurate \textsc{Licorice} power spectra given $k,z$, and a set of the four astrophysical parameters. During training, \textsc{LorEMU} reaches between $\sim5\%$ and $10\%$ relative mean squared error averaged over the dataset. We then perform Bayesian inference using \textsc{LorEMU} and MCMC on a power spectrum generated using \textsc{Licorice}, to which a noise realization corresponding to 100h of SKA observations was added. Because different random weight initializations affect the result of the network training, nine different versions of  \textsc{LorEMU} were trained, and training variance was included into the inference likelihood. By averaging the outputs of the different networks during inference, we obtain accurate posterior distributions. We find that this approach produces a posterior with very little bias by performing inference on a noiseless simulated signal.
Finally, we apply the same inference pipeline to the most recent HERA upper limits and obtain constraints on lower values of $f_X$, indicating that our Universe is unlikely to have followed a cold reionization scenario.

The new implementation of \textsc{Licorice} suffers from some limitations. While star formation in resolved halos is sensitive to the local environment through the dynamics of the gas, unresolved halos of a given mass have a similar SFR, which only varies if they are located in particles with different stellar fractions. Moreover, the CMF formalism provides the cosmic average of the halo population in a given region with fixed overdensity. In reality, such regions may present fluctuations around this average. A future improvement will be to include stochasticity in the unresolved SFR, both from the halo population fluctuations and from the star formation efficiency of each unresolved halo. The latter will have to be parameterized with one last additional astrophysical parameter. Future versions of the LoReLi database will include both a denser sampling of the parameter space and variations of additional astrophysical parameters.

In this work, we have only begun to tap the potential of the LoReLi database. As we are modeling the 21 cm signal using full nonlinear dynamics and 3D radiative transfer, we can hope that the non-Gaussian properties of the signal are well accounted for. It will be relevant to explore inference based on non-Gaussian summary statistics of the signal. An explicit likelihood is usually not available in this case. One solution is to choose summary statistics that, while encoding non-Gaussian properties of the signal, have themselves near-Gaussian behaviors (due to the central limit theorem) and model the likelihood as a multivariate Gaussian. Another promising avenue is to use simulation-based inference (e.g., \cite{Zhao2022}). We will be working toward both goals.

\begin{acknowledgements}
     This project was provided with computer and storage resources by GENCI at
TGCC thanks to the grant 2023-A0130413759 on the supercomputer
Joliot Curie's ROME
partition.
This study was granted access to the HPC resources of MesoPSL financed by the Région
Île-de-France and the project EquipMeso (reference ANR-10-EQPX-29-01) of the programme
Investissements d’Avenir supervised by the Agence Nationale pour la Recherche.
\end{acknowledgements}

%
%

\bibliographystyle{aa} 
\bibliography{My_Collection} 



\end{document}